\pdfoutput=1
\documentclass[floatfix,amsmath,amssymb,
    aps, pra,
    twocolumn,mathbbm, superscriptaddress,nofootinbib]{revtex4-2}
\usepackage{graphicx}
\usepackage{physics}
\usepackage{siunitx}
\usepackage{mathtools}
\usepackage{float}
\usepackage{bm}
\usepackage{bbm}
\usepackage[dvipsnames, table]{xcolor}
\usepackage[colorlinks=true,citecolor=Cerulean,linkcolor=RubineRed,urlcolor=Cerulean]{hyperref}
\usepackage[T1]{fontenc}
\usepackage{letltxmacro}

\LetLtxMacro{\originaleqref}{\eqref}
\renewcommand{\eqref}{Eq.~\ref}

\newcommand{\up}{{\ket{\uparrow}}}
\newcommand{\down}{{\ket{\downarrow}}}
\newcommand{\Hsoc}{{H_{\text{SOC}}}}
\newcommand{\wpwidth}{\zeta}
\newcommand{\fleft}{\mathopen{}\mathclose\bgroup\left}
\newcommand{\fright}{\aftergroup\egroup\right}

\begin{document}
\title{Interface-Induced Conservation of Momentum Leads to Chiral-Induced Spin Selectivity}

\author{Clemens Vittmann}
\affiliation{Institut f\"ur Theoretische Physik und IQST, Albert-Einstein-Allee 11, Universit\"at Ulm, D-89081 Ulm, Germany}
\author{R. Kevin Kessing}
\affiliation{Institut f\"ur Theoretische Physik und IQST, Albert-Einstein-Allee 11, Universit\"at Ulm, D-89081 Ulm, Germany}
\author{James Lim}
\affiliation{Institut f\"ur Theoretische Physik und IQST, Albert-Einstein-Allee 11, Universit\"at Ulm, D-89081 Ulm, Germany}
\author{Susana F. Huelga}
\affiliation{Institut f\"ur Theoretische Physik und IQST, Albert-Einstein-Allee 11, Universit\"at Ulm, D-89081 Ulm, Germany}
\author{Martin B. Plenio}
\email{martin.plenio@uni-ulm.de}
\affiliation{Institut f\"ur Theoretische Physik und IQST, Albert-Einstein-Allee 11, Universit\"at Ulm, D-89081 Ulm, Germany}

\begin{abstract}
We study the non-equilibrium dynamics of electron transmission from a straight waveguide to a helix with spin--orbit coupling.
Transmission is found to be spin-selective and can lead to large spin polarizations of the itinerant electrons.
The degree of spin selectivity depends on the width of the interface region, and no polarization is found for single-point couplings.
We show that this is due to momentum conservation conditions arising from extended interfaces.
We therefore identify interface structure and conservation of momentum as crucial ingredients for chiral-induced spin selectivity, and confirm that this mechanism is robust against static disorder.
\end{abstract}

\maketitle

Chiral-induced spin selectivity (CISS)~\cite{aiello2020chiralitybased,naaman_spintronics_2015} is an intriguing phenomenon wherein the chirality of organic molecules such as DNA~\cite{Goehler894}, oligopeptides~\cite{kettner_spin_2015}, bacteriorhodopsine~\cite{mishra_spin-dependent_2013} or photosystem~I~\cite{carmeli_spin_2014} induces notable spin dependence on various electronic processes. 
Experiments have demonstrated that unpolarized photoelectrons 
can become highly spin-polarized when transmitted through (mono\nobreakdash-)layers of helical molecules, with the molecule's handedness deciding the sign of the polarization~\cite{EnantiospecificSpinPol_2014, kettner_2018}. 
The CISS effect has also been observed for bound electron transport in the form of spin-dependent currents when helical molecules are placed between a substrate and an atomic force microscope (AFM) tip~\cite{dna_oligomers_2011}. 
Naturally, such phenomena have attracted much interest in the field of spintronics that uses the electron spin to process information in nanodevices~\cite{bostick_2018, dor_chiral-based_2013, Michaeli_2017, mondal_spin-dependent_2016, yang_2020, Chiesa2021}. 
More recently, the relations between electron spin, enantioselectivity and chemical reactivity in chiral molecules has also been the subject of increased attention~\cite{kumar_chirality-induced_2017, banerhee-ghosh_2018, metzger_electron_2020, dianat_2020, kapon_evidence_2021}.
Though it is apparent that the experimentally observed spin polarization is related to the spin--orbit coupling (SOC) experienced by the electrons moving through the helical or chiral structures, a consensus on the microscopic mechanism has not yet been reached~\cite{Overview2021}. 

\begin{figure}[H]%
	\includegraphics[width=0.47\textwidth]{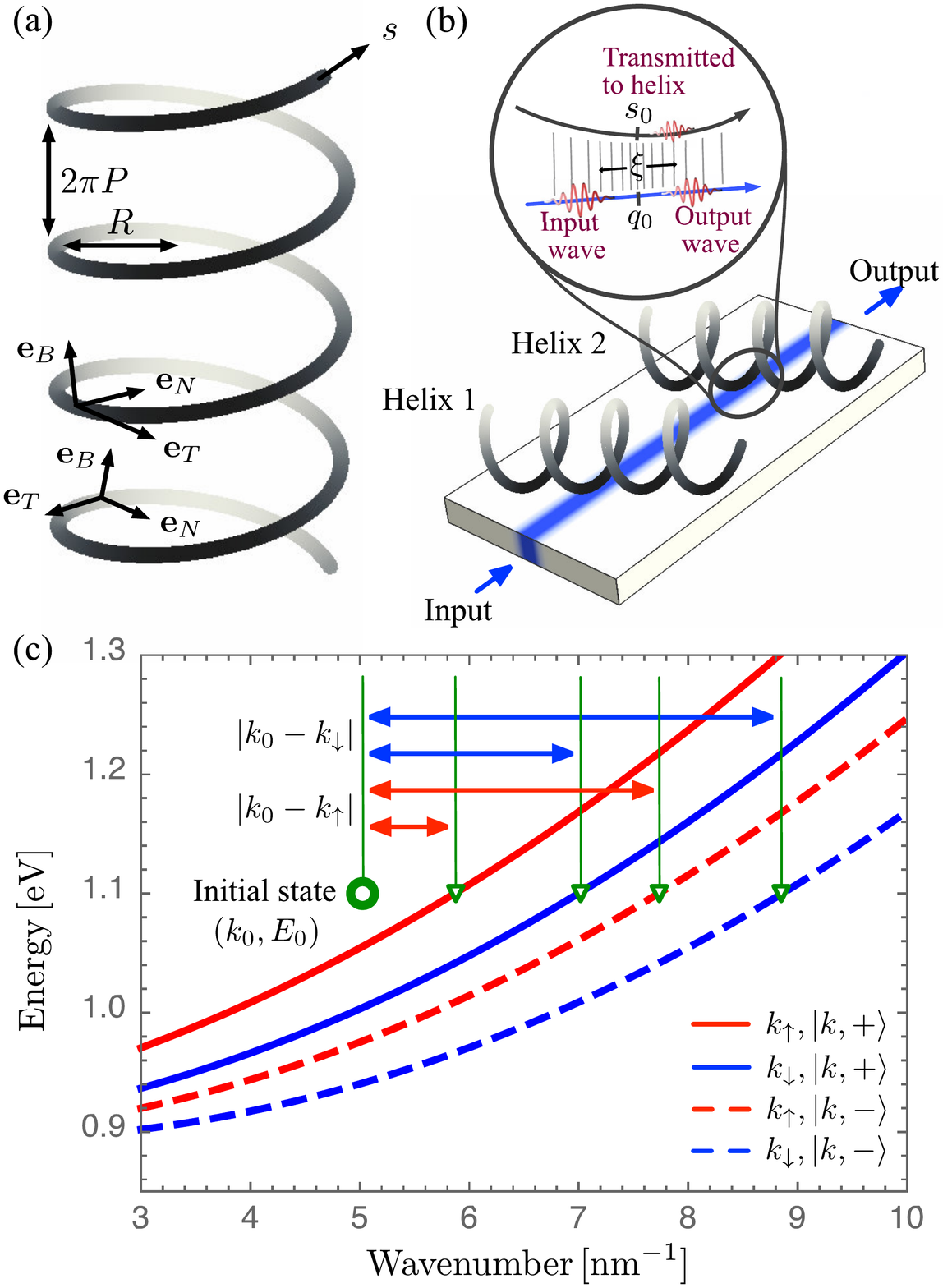}%
	\caption{(a)~Helix-shaped waveguide with radius $R$ and pitch $2\pi P$. At each point, a coordinate system is defined by the normal ($\mathbf{e}_N$), bi-normal ($\mathbf{e}_B$) and tangential ($\mathbf{e}_T$) vectors. 
	(b)~Schematic representation of multiple scattering events of a electron wavepacket traveling along a SOC-free waveguide. The scattering occurs over a region of non-zero length $\xi$, leading to a transmitted wavepacket in the helical waveguide and a remaining wavepacket in the free waveguide. 
	(c)~Spin-dependent dispersion relations of a helical waveguide. The initial wavenumber $k_0$ and energy $E_0$ of an input state are marked by a green circle. The difference between the input wavenumber $k_0$ and spin-dependent wavenumbers $k_\uparrow$ and $k_\downarrow$ of the $E_0$-energetically resonant eigenstates of the helical waveguide are highlighted by red and blue arrows, respectively.}%
	\label{Fig1}%
\end{figure}

In this work, we present a simple and natural model exhibiting CISS whose underlying mechanism can be rigorously understood using intuitive physical concepts:
Electrons initially traveling along a SOC-free straight waveguide approach a helical waveguide (Fig.~\ref{Fig1}(a)) with SOC to which they can tunnel, resulting in an effective coupling region of width $\xi$ (Fig.~\ref{Fig1}(b)).
We show that an initially spin-unpolarized electron wavepacket naturally becomes polarized after coming into proximity with the helix \emph{if} the interaction region is not a single point, i.e., if $\xi > 0$.
This process can be repeated if the electron then comes into contact with further helices, amplifying the effect.
We demonstrate that high spin polarization $\sim \SI{50}{\percent}$ can be achieved for moderate SOC strengths $\hbar\alpha\sim \SI{1}{\meV\nm}$, despite the helical waveguide being modeled as a single one-dimensional channel without -- contrary to previous theoretical studies -- intramolecular
\noindent multi-channel structures. 
However, when the SOC-free and helical waveguides are coupled only at a single point ($\xi \to 0$), the outgoing state remains unpolarized.
As we will show in this work, these results can be rationalized in terms of spin-dependent inter-waveguide transmission probabilities, as an extended interaction region $\xi > 0$ prescribes (partial) conservation of momentum, resulting in spin-selective transmissions due to spin--orbit coupling within the helix.
Therefore, our work highlights the importance of interface modeling in CISS.

As shown in Fig.~\ref{Fig1}(a), each helical molecule is characterized by its radius $R$ and pitch $2\pi P$; we use the geometrical parameters of DNA, $R=\SI{0.7}{\nm}$ and $2\pi P=\SI{3.4}{\nm}$~\cite{naaman_chiral-induced_2012}, throughout.
As a model for an electron traveling through the potential energy landscape of a helical molecule, we assume that an electron wavepacket propagates solely in the tangential direction $\mathbf{e}_T$ of a helical path due to a trapping potential acting in the normal $\mathbf{e}_N$ and binormal $\mathbf{e}_B$ directions. 
The electron's dynamics can then be described by an effective one-dimensional quantum state in the form $\psi_{\uparrow}(s,t)\ket{\uparrow}+\psi_{\downarrow}(s,t)\ket{\downarrow}$, where $\ket{\uparrow}$ and $\ket{\downarrow}$ are the eigenstates of the Pauli matrix $\sigma_z$, $t$ is the time coordinate and $s$ represents the position on the helical path, as shown in Fig.~\ref{Fig1}(a). 
As in previous CISS models~\cite{Cuniberti2012, guo_spin-dependent_2014, GutierrezCunibertiEffective, naaman_chiral_2019, Naaman_Michaeli_2019, GhazaryanAnalyticModel2020}, we assume that the electric field induced by the molecule is helical-symmetric and that the field on the helical path is oriented in the normal direction, $\mathbf{E}=E\mathbf{e}_N(s)$ with constant $E$, where the normal unit vector $\mathbf{e}_N(s)$ depends on the position $s$.
The dynamics of the electron wavepacket is governed by an effective Hamiltonian~\cite{ortix_quantum_2015} of the form
\begin{align} 
    H_h=\frac{{p_s}^2}{2m_h}+U_h+ \frac{\alpha}{2}\acomm{\bm{\sigma}\cdot \mathbf{e}_B(s)}{p_s},\label{eq:H}
\end{align}
where $p_s=-i\hbar\partial_s$ is the momentum operator in the tangential direction $\mathbf{e}_T(s)$, $m_h=10\,m_e$ is the effective electron mass~\cite{EffectiveMass1, EffectiveMass2} with $m_e$ its rest mass, and $U_h$ is a constant potential energy shift. The last term in \eqref{eq:H} represents the spin--orbit interaction, which is generally described by $\Hsoc = -\bm{\sigma}\cdot(\bm{\alpha}\times \mathbf{p})$,
where $\bm{\alpha}$ is proportional to $\mathbf{E}$ and quantifies the SOC strength via its magnitude $\alpha$. Thus, since we assume $\mathbf{E}$ parallel to $\mathbf{e}_N(s)$, we have $\bm{\alpha} = \alpha \mathbf{e}_N(s)$ and $\Hsoc$ as shown in \eqref{eq:H}.

As shown in the SI, the eigenstates as a function of the helical position coordinate $s$ can be expressed as
\begin{equation}
    \braket{s}{k,\pm} =\frac{A_{\pm} e^{ik_\uparrow s}\ket{\uparrow}+e^{ik_\downarrow s}\ket{\downarrow}}{\sqrt{2\pi\left(1+\abs{A_\pm}^2\right)}},\label{eq:eigenstates}
\end{equation}
with $D=\sqrt{R^2+P^2}$, where the spin-dependent wavenumbers are given by $k_\uparrow = k - (2D)^{-1}$ and $k_\downarrow = k+(2D)^{-1}$, 
and $A_+$ and $A_-$ are $k$-independent coefficients.
This shows that there are two distinct groups of eigenstates, $\{\ket{k,+}\}$ and $\{\ket{k,-}\}$, with different energies and opposite spin polarizations $P_\pm = \expval{\sigma_z} =(\abs{A_\pm}^2-1)/(\abs{A_\pm}^2+1)$ which fulfill $P_+ = -P_-$. 
In Fig.~\ref{Fig1}(c), the dispersion relation between energy $E(k,+)$ and wavenumber $k_\uparrow$ (or $k_\downarrow$) of the eigenstates $\{\ket{k,+}\}$ is shown in a red (or blue) solid line, where $U_h=\SI{0.9}{\eV}$ and $\hbar \alpha=\SI{10}{\meV\nm}$ are considered as an example. 
These $k_\uparrow$ and $k_\downarrow$ curves differ due to the correlations between spin states and wavenumbers induced by spin--orbit interaction, as shown in  \eqref{eq:eigenstates}.
Similarly, the two components of the dispersion relation of the $\{\ket{k,-}\}$ eigenstates are shown in red and blue dashed lines, respectively, in Fig.~\ref{Fig1}(c).
These results demonstrate that when an electron with well-defined energy $E_0$ and wavenumber $k_0$ enters a helical waveguide in such a way that the energy of an electron wavepacket is conserved (such that $E_0=E(k,\pm)$), the wavenumber $k_\uparrow$ or $k_\downarrow$ of the helical eigenstate $\ket{k,\pm}$ will generally differ from $k_0$ of the input state. 
Importantly, the difference between wavenumbers can be smaller or larger depending on the incoming state's polarization.
In the case shown in Fig.~\ref{Fig1}(c), the difference is smaller for a spin-up state than for a spin-down state, $\abs{k_0-k_\uparrow} < \abs{k_0-k_\downarrow}$, as highlighted by red and blue arrows. 
Therefore, if the transition probability favors smaller differences between wavenumbers --  that is, if momentum conservation is (at least partially) enforced -- then a $\up$ state will transition to the helix more readily than a $\down$ state.
As we will show shortly, this partial momentum conservation does indeed arise when the interfacial coupling is extended over a region rather than being point-like, thus constituting the centerpiece of our model.
Finally, it follows from these preliminaries that an initially unpolarized mixed state 
will lose $\up$ population to the helix more quickly than $\down$: This results in a net negative spin polarization of the remaining electronic density that has not been transmitted to the helix,
an effect which can be further amplified by repeated interactions with other helices.

To quantitatively model the relation between interfacial structure and momentum-conservation assisted spin filtering, we introduce a second one-dimensional waveguide. 
This second waveguide shall be governed by a free-space Hamiltonian in the form $H_f=(2m_e)^{-1}{p_q}^{2}$, where $p_{q}=-i\hbar\partial_{q}$ is the momentum operator along a separate coordinate $q$, as shown in the inset of Fig.~\ref{Fig1}(b). 
The eigenstates of $H_f$ are spin-independent and given by $\braket{q}{k}=(2\pi)^{-1/2}e^{ik q}$ with wavenumbers $k$ (and $\braket{s}{k} = 0$). 
We call this the \emph{free waveguide} or free-space channel, as opposed to the helical waveguide.
To model an inter-waveguide scattering process occurring over an extended region, we consider coupling between free (coordinate $q$) and helical waveguides ($s$) of the form 
\begin{equation}
V(s, q) =  V_0 \exp(-\frac{(s-s_0)^2}{2\xi^2}) \delta\fleft((s-s_0) - (q-q_0)\fright)
\label{eq:Vsq}
\end{equation}
where $s$ and $q$ denote the position on the helical and free waveguide, respectively and $\delta(x)$ is the Dirac delta, which signifies that each position is coupled to exactly one position on the opposite waveguide.
This $V(s,q)$ implies that when an electron wavepacket propagates in the free waveguide, the phase information of the input state, encoded as a function of the free-space coordinate $q$, is transferred to the helical waveguide over an extended width $\xi$ of the interface centered at $(s,q)=(s_0,q_0)$, as shown schematically in the inset to Fig.~\ref{Fig1}(b).
When the input state of the free-space channel is a $\ket{\uparrow}$ state with wavenumber $k_0$, the coupling to a helical eigenstate $\ket{k,\pm}$ is given by 
\begin{equation}
\abs{\!\mel{k,{\pm}}{V}{k_0,\uparrow}}^2 = \left(\frac{{V_0}^2 \xi^2}{2\pi} \frac{\abs{A_{\pm}}^2}{1+\abs{A_\pm}^2} \right) e^{-\xi^2 \left(\Delta k\right)^2}
\label{eq:mel}
\end{equation}
with $\Delta k = k_\uparrow-k_0$. 
We see that the coupling strength is stronger if the eigenstate $\ket{k,\pm}$ has a larger $\ket{\uparrow}$ component amplitude $\abs{A_\pm}$  (cf.~\eqref{eq:eigenstates}). 
However, more importantly, as the width $\xi$ of the interface increases, the effective coupling strength becomes more sensitive to the momentum mismatch $\hbar \Delta k$, and as a result transfer occurs only if the wavenumber is conserved, $k_\uparrow \approx k_0$. 
On the other hand, when the interface is very narrow, $\xi \to 0$, the effective coupling strength becomes independent of $\Delta k$. 
In this case, the only information provided to a helical channel is the time-dependent amplitude of the input state at a single-point contact with the free waveguide.
This makes scattering probabilities independent of changes in momentum, similar to the textbook problem of one-dimensional scattering at a step potential~\cite{Shankar}.

\begin{figure}
	\includegraphics[width=0.48\textwidth]{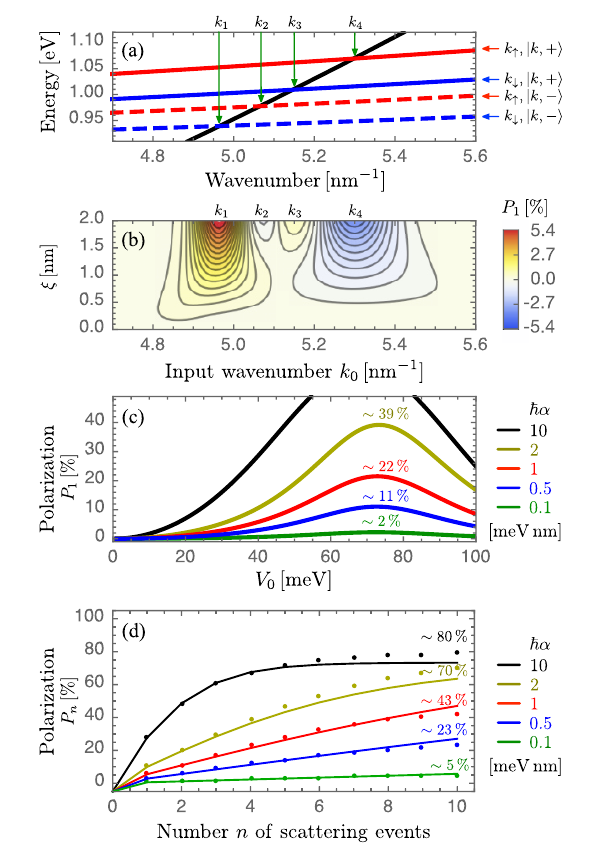}
	\caption{(a)~Spin-dependent dispersion relations of a helical molecule (red and blue, cf.~Fig.~\ref{Fig1}(c)), and the free-space dispersion relation (black), for $U_h = \SI{0.9}{\eV}$ and $\hbar\alpha = \SI{10}{\meV\nm}$. 
	The $k_{1,2,3,4}$ of the four crossing points are highlighted by green arrows. 
	(b)~Spin polarization $P_1$ after a single scattering event as a function of input wavenumber $k_0$ and the interaction region width $\xi$ (cf.~Fig.~\ref{Fig1}(b)). $U_h$ and $\alpha$ as in (a), with $V_0=\SI{10}{\meV}$. 
	(c)~$P_1$ as a function of the interfacial coupling strength $V_0$ for $\xi = \SI{1}{\nm}$. 
	We use $k_0=\SI{5.075}{\per\nm}$ except for $\hbar\alpha = \SI{10}{\meV\nm}$, where \SI{4.975}{\per\nm} is considered due to the $\alpha$-dependence of the crossing points.
	(d)~Spin polarization $P_n$ after $n$ scattering events with parameters as given in (c) and $V_0 = \SI{40}{\meV}$. Dots are numerical calculations, solid lines are transfer-matrix-based extrapolations.}
	\label{Fig2}
\end{figure}
To demonstrate that a non-zero length $\xi$ of the interface can induce the momentum-conservation assisted spin filtering, we consider a Gaussian initial electron state in the form 
\begin{equation}
    \braket{q}{\psi_i^\sigma(t=0)} = 
    \frac{1}{\pi^\frac{1}{4} \wpwidth^{\frac{1}{2}}} 
    \exp(-\frac{(q - q_i)^2}{2 \wpwidth^2 } + ik_0 q) 
    \ket{\sigma},
    \label{eq:incoming_state}
\end{equation}
with $\sigma  \in \{\uparrow, \downarrow\}$,
and with an initial center position $q_i$
such that the amplitude at the interface around $q_0$ is negligible. 
This is a superposition of the eigenstates $\ket{k}$ of the free-space Hamiltonian $H_f$ such that $\abs{\braket{k}{\psi_i^\sigma}}^2 \propto e^{- (k - k_0)^2 \wpwidth^2}$, which is centered at $\expval{k} = \expval{p}/\hbar = k_0$ with an uncertainty $\sigma_k = \left(\sqrt{2} \wpwidth\right)^{-1}$ (we use $\zeta = 100/k_0$ throughout).
We can then apply time-dependent perturbation theory to first order in $V$ (\eqref{eq:Vsq}), similar to Fermi's Golden Rule, to obtain the probability that an incoming Gaussian state $\ket{\psi_i^\sigma}$ of polarization $\sigma$, \eqref{eq:incoming_state}, has transitioned to a helix eigenstate $\ket{k_\text{out}, \pm}$, \eqref{eq:eigenstates}, in the long-time limit $t\to\infty$ (see SI for  further details):
\begin{multline}
\lim_{t\to\infty} \abs{\braket{k_\text{out}, \pm}{\psi_i^\sigma(t)}}^2 \\
\propto 
\abs{\int_\mathbb{R}
e^{- \frac{\xi^2\left(\Delta k\right)^2}{2} - \frac{\wpwidth^2 \left(k - k_0\right)^2}{2} - i \phi(k)} \,
\delta\fleft(\omega(k)\fright) \dd{k}}^2,
\label{eq:PT}
\end{multline}
where $\Delta k= k_\sigma - k$ and $\omega(k) = \hbar k^2/(2m_e) - E(k_\text{out}, \pm)/\hbar$, 
and $\phi(k) = (q_i - q_0) k$ is a (largely irrelevant) phase.
We see that in the long-time limit, energy is rigorously conserved during transitions, but momentum is only loosely conserved according to a Gaussian law, as predicted by \eqref{eq:mel}. For small $\xi$ and/or $V_0$, such that the influence of $V(s,q)$ remains perturbative, this simple equation can be used to accurately describe the transition probabilities and therefore the spin polarizations of our model.
We quantify the spin polarization arising from an initially unpolarized mixed state $\rho_0 = \frac{1}{2}\left[\dyad{\psi_i^\uparrow} + \dyad{\psi_i^\downarrow}\right]$ with $\ket{\psi_i^\sigma}$ from \eqref{eq:incoming_state} as 
\[P_1 = \Tr{\left[\mathcal{P}_f \rho_1 \mathcal{P}_f^\dag\right] \sigma_z}/\Tr{\mathcal{P}_f \rho_1  \mathcal{P}_f^\dag}\]
where $\rho_1$ is the state $\rho_0$ after passing through an interaction region with a helix, and $\mathcal{P}_f$ is the projector onto the free-waveguide component.
The polarization $P_n$ after $n$ repeated interactions with separate helices is defined analogously using the state $\rho_n$, 
which is the result of scattering the state $\mathcal{P}_f \rho_{n-1} \mathcal{P}_f^\dag$. 

To expand our analysis to the non-perturbative regime, we calculate numerically the dynamics of the wavepacket using a
standard finite-difference scheme
(see SI), which further shows good agreement with \eqref{eq:PT} within the perturbative regime.
We illustrate the rich physics of our model by demonstrating the effect of varying the interfacial length $\xi$ and the incoming momentum $\hbar k_0$:
In Fig.~\ref{Fig2}(a), the dispersion relations of the spin components of the helical eigenstates $\ket{k,\pm}$ are shown in red and blue, in addition to that of the free-space eigenstates, shown in black. The dispersion curves are crossed at four different wavenumbers $k_{1,2,3,4}$, highlighted by green arrows. 
In Fig.~\ref{Fig2}(b), where $\hbar\alpha = \SI{10}{\meV\nm}$ and $V_0 = \SI{10}{\meV}$, the spin polarization $P_1$ after a single scattering event is shown as a function of the input wavenumber $k_0$ and the width $\xi$ of the Gaussian coupling spectrum of the interface. As expected from the above discussion, the spin polarization is zero for $\xi=0$ (single-point coupling) and the spin polarization effect is enhanced as $\xi$ increases.
At the crossing point of the momenta $k_{1,3}$ ($k_{2,4}$), the spin polarization becomes maximally positive (negative), as these momenta correspond to the $\down$ ($\up$) component of the helical eigenstates, causing an incoming $\down$ ($\up$) state to be transmitted to the helix with higher probability, as predicted by \eqref{eq:PT}. 
In Fig.~\ref{Fig2}(c), where $\xi = \SI{1}{\nm}$ and $k_0\approx \SI{5}{\per\nm}$, the spin polarization is displayed as a function of the interfacial coupling strength $V_0$ for different spin--orbit coupling strengths from \SIrange{0.1}{10}{\meV\nm}. 
It is notable that even for weak spin--orbit coupling $\hbar\alpha = \SI{0.1}{\meV\nm}$ and $\SI{1}{\meV\nm}$, the spin polarization reaches values up to $\sim \SI{2}{\percent}$ and $\sim \SI{22}{\percent}$, respectively. 
Figure~\ref{Fig2}(d) shows the results of numerical simulations (shown as dots) with $V_0=\SI{40}{\meV}$ demonstrating that the spin polarization is further enhanced when the input wavepacket is scattered multiple times by independent helices (cf.~Fig.~\ref{Fig1}(b)). 
For example, in the case of weak spin--orbit coupling strength $\hbar\alpha = \SI{1}{\meV\nm}$, the spin polarization is increased from $P_1\approx \SI{6}{\percent}$ to $P_{10}\approx \SI{43}{\percent}$ after 10 consecutive scattering events. 
Furthermore, using a transfer-matrix-like method, we can estimate the $n$-scattering polarization $P_n$ based only on the single-scattering data, which also shows that $\abs{P_\infty} < \SI{100}{\percent}$ is caused by spin-flip back-transfer from the helical to straight waveguide (solid lines in Fig.~\ref{Fig2}(d), see SI for details). 
These results demonstrate that high spin polarization can be obtained from initially unpolarized electrons when scattering occurs over an extended region of a helical molecule even for weak intrinsic SOC strength ($\hbar\alpha\lesssim \SI{1}{\meV\nm}$).

Notice that the vanishing spin polarization in the case of single-point couplings is in line with the well-known fact that in one-dimensional theories, spin--orbit coupling can be removed by a unitary transformation and hence -- on its own -- cannot account for spin-filtering~\cite{UtsumiComment}. Previous studies have therefore claimed that multiple channels such as additional helix strands~\cite{Guo_2012,sierra_spin-polarized_2019} or orbitals~\cite{GutierrezCunibertiEffective,Utsumi2020} are required to observe spin polarization.  
As shown in the SI, the extended nature of the interface coupling prevents spin-dependence in our model from being removed by a unitary transformation and no intramolecular multi-channel structure need be employed. 

\begin{figure}
	\includegraphics[width=0.47\textwidth]{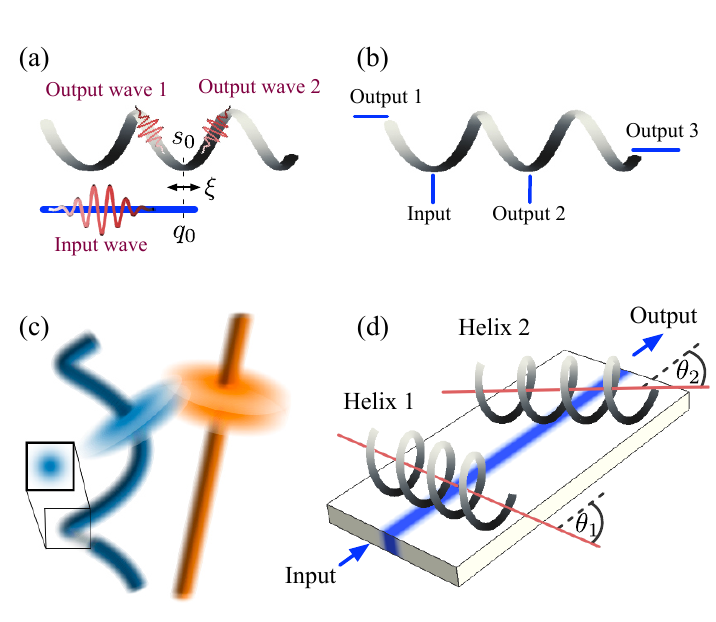}
	\caption{(a)~Three-terminal setup where the tip of an electrode is coupled to a single helix over an extended region. An unpolarized input wavepacket is transmitted to the helix and results in oppositely polarized wavepackets traveling in opposite directions. 
	(b)~In a general $N$-terminal setup where all interfaces are point contacts, no spin polarization can be observed.
	(c)~To incorporate geometrical effects, an inter-waveguide coupling matrix is calculated from the overlap between Gaussian cross-section slices: One slice--wedge pair is shown exploded.
	(d)~Schematic representation of multiple scattering events with varying orientations of the helix axes. Helix 2 is shown in the locally parallel configuration $\theta_2 \approx \SI{52}{\degree}$, whereas helix 1 is not.}
	\label{Fig3}
\end{figure}
The mechanism of momentum conservation and spin-dependent transmission can also be used to understand spin-polarized transport through a helical waveguide attached to leads: Figure~\ref{Fig3}(a) illustrates a three-terminal setup with an electrode tip coupled to a helix over a length $\xi$. A wavepacket transmitted to the helix splits into oppositely polarized wavepackets propagating in opposite directions inside the helix. In this case -- as well as in any general $N$-terminal setup as depicted in Fig.~\ref{Fig3}(b) -- average spin polarization vanishes if all couplings are point contacts.

Lastly, the coupling function $V(s,q)$ of \eqref{eq:Vsq} may be a simplification of the inter-waveguide coupling, as it does not  directly depend on the 3D arrangement of the system. 
Therefore, to further validate our interface-based mechanism and test its robustness against disorder, we have calculated dynamics using a more elaborate coupling function $V(s,q)$ based on the 3D overlap between Gaussian cross-section wavefunctions occupying the straight and helical systems, as illustrated in Fig.~\ref{Fig3}(c). 
This more advanced coupling model naturally depends on the angle and distance between the helix and the straight waveguide, and it is also no longer diagonal in the sense that $V = 0$ does not necessarily hold for $s-s_0 \neq q-q_0$. 
Using this geometry-dependent model, we examine the influence of variations of the angle to the helix, illustrated in Fig.~\ref{Fig3}(d).
The simpler coupling model of \eqref{eq:Vsq} is closest to the geometrical configuration in which the tangent vectors of the two waveguides are parallel at their point of least separation, given by $\theta = \arccos(P/D) \approx \SI{52}{\degree}$:
We find that variations of $\SI{\pm 20}{\degree}$ about this angle have little impact on the spin selectivity (shown in SI).
We therefore conclude that our model is sufficiently robust against static disorder to be measurable in experimental setups similar to those proposed here.

In conclusion, using a simple model consisting of a straight electron waveguide coupled to a helical waveguide, we have shown that electron wavepackets can become strongly polarized due to spin-dependent transmission when the inter-waveguide coupling is extended over a region of non-zero length. A multichannel structure of the helical molecule is not required.
These effects are robust against static disorder and can be enhanced by consecutive helix interactions.
By quantitatively investigating the relationship between the coupling region width and the spin polarization, we identified interface-induced (partial) conservation of momentum as a crucial ingredient to this manifestation of CISS, pointing us to an intuitive and well-established principle to guide further studies of spin selectivity.
This complements existing work on interface design in CISS experiments~\cite{GuoContactEffects2014} by providing a directly deducible and simple mechanism without the need to invoke further phenomena such as dephasing.
This mechanism is found to arise naturally in the continuum case, and further analysis may reveal how our findings are changed when considering a discretized tight-binding case or how the principle of momentum conservation could be applied to setups involving transport through helical molecules.
Furthermore, our findings can be investigated experimentally using a setup as shown in Fig.~\ref{Fig3}(d) with waveguides or leads of varying thickness representing different interface widths.

\begin{acknowledgments}
This work was supported by the ERC Synergy grant HyperQ (grant no.\ 856432).
The authors acknowledge support by the state of Baden-Württemberg through bwHPC and the German Research Foundation (DFG) through grant no.\ INST 40/575-1 FUGG (JUSTUS~2 cluster).
\end{acknowledgments}

\bibliography{bibliography}

\begin{thebibliography}{37}%
\makeatletter
\providecommand \@ifxundefined [1]{%
 \@ifx{#1\undefined}
}%
\providecommand \@ifnum [1]{%
 \ifnum #1\expandafter \@firstoftwo
 \else \expandafter \@secondoftwo
 \fi
}%
\providecommand \@ifx [1]{%
 \ifx #1\expandafter \@firstoftwo
 \else \expandafter \@secondoftwo
 \fi
}%
\providecommand \natexlab [1]{#1}%
\providecommand \enquote  [1]{``#1''}%
\providecommand \bibnamefont  [1]{#1}%
\providecommand \bibfnamefont [1]{#1}%
\providecommand \citenamefont [1]{#1}%
\providecommand \href@noop [0]{\@secondoftwo}%
\providecommand \href [0]{\begingroup \@sanitize@url \@href}%
\providecommand \@href[1]{\@@startlink{#1}\@@href}%
\providecommand \@@href[1]{\endgroup#1\@@endlink}%
\providecommand \@sanitize@url [0]{\catcode `\\12\catcode `\$12\catcode
  `\&12\catcode `\#12\catcode `\^12\catcode `\_12\catcode `\%12\relax}%
\providecommand \@@startlink[1]{}%
\providecommand \@@endlink[0]{}%
\providecommand \url  [0]{\begingroup\@sanitize@url \@url }%
\providecommand \@url [1]{\endgroup\@href {#1}{\urlprefix }}%
\providecommand \urlprefix  [0]{URL }%
\providecommand \Eprint [0]{\href }%
\providecommand \doibase [0]{https://doi.org/}%
\providecommand \selectlanguage [0]{\@gobble}%
\providecommand \bibinfo  [0]{\@secondoftwo}%
\providecommand \bibfield  [0]{\@secondoftwo}%
\providecommand \translation [1]{[#1]}%
\providecommand \BibitemOpen [0]{}%
\providecommand \bibitemStop [0]{}%
\providecommand \bibitemNoStop [0]{.\EOS\space}%
\providecommand \EOS [0]{\spacefactor3000\relax}%
\providecommand \BibitemShut  [1]{\csname bibitem#1\endcsname}%
\let\auto@bib@innerbib\@empty
\bibitem [{\citenamefont {Aiello}\ \emph {et~al.}(2020)\citenamefont {Aiello},
  \citenamefont {Abbas}, \citenamefont {Abendroth}, \citenamefont {Banerjee},
  \citenamefont {Beratan}, \citenamefont {Belling}, \citenamefont {Berche},
  \citenamefont {Botana}, \citenamefont {Caram}, \citenamefont {Celardo},
  \citenamefont {Cuniberti}, \citenamefont {Dianat}, \citenamefont {Guo},
  \citenamefont {Gutierrez}, \citenamefont {Herrmann}, \citenamefont {Hihath},
  \citenamefont {Kale}, \citenamefont {Kurian}, \citenamefont {Lai},
  \citenamefont {Medina}, \citenamefont {Mujica}, \citenamefont {Naaman},
  \citenamefont {Noormandipour}, \citenamefont {Palma}, \citenamefont
  {Paltiel}, \citenamefont {Petuskey}, \citenamefont {Ribeiro-Silva},
  \citenamefont {Stemer}, \citenamefont {Valdes-Curiel}, \citenamefont
  {Varela}, \citenamefont {Waldeck}, \citenamefont {Weiss}, \citenamefont
  {Zacharias},\ and\ \citenamefont {Wang}}]{aiello2020chiralitybased}%
  \BibitemOpen
  \bibfield  {author} {\bibinfo {author} {\bibfnamefont {C.~D.}\ \bibnamefont
  {Aiello}}, \bibinfo {author} {\bibfnamefont {M.}~\bibnamefont {Abbas}},
  \bibinfo {author} {\bibfnamefont {J.}~\bibnamefont {Abendroth}}, \bibinfo
  {author} {\bibfnamefont {A.~S.}\ \bibnamefont {Banerjee}}, \bibinfo {author}
  {\bibfnamefont {D.}~\bibnamefont {Beratan}}, \bibinfo {author} {\bibfnamefont
  {J.}~\bibnamefont {Belling}}, \bibinfo {author} {\bibfnamefont
  {B.}~\bibnamefont {Berche}}, \bibinfo {author} {\bibfnamefont
  {A.}~\bibnamefont {Botana}}, \bibinfo {author} {\bibfnamefont {J.~R.}\
  \bibnamefont {Caram}}, \bibinfo {author} {\bibfnamefont {L.}~\bibnamefont
  {Celardo}}, \bibinfo {author} {\bibfnamefont {G.}~\bibnamefont {Cuniberti}},
  \bibinfo {author} {\bibfnamefont {A.}~\bibnamefont {Dianat}}, \bibinfo
  {author} {\bibfnamefont {Y.}~\bibnamefont {Guo}}, \bibinfo {author}
  {\bibfnamefont {R.}~\bibnamefont {Gutierrez}}, \bibinfo {author}
  {\bibfnamefont {C.}~\bibnamefont {Herrmann}}, \bibinfo {author}
  {\bibfnamefont {J.}~\bibnamefont {Hihath}}, \bibinfo {author} {\bibfnamefont
  {S.}~\bibnamefont {Kale}}, \bibinfo {author} {\bibfnamefont {P.}~\bibnamefont
  {Kurian}}, \bibinfo {author} {\bibfnamefont {Y.-C.}\ \bibnamefont {Lai}},
  \bibinfo {author} {\bibfnamefont {E.}~\bibnamefont {Medina}}, \bibinfo
  {author} {\bibfnamefont {V.}~\bibnamefont {Mujica}}, \bibinfo {author}
  {\bibfnamefont {R.}~\bibnamefont {Naaman}}, \bibinfo {author} {\bibfnamefont
  {M.}~\bibnamefont {Noormandipour}}, \bibinfo {author} {\bibfnamefont
  {J.}~\bibnamefont {Palma}}, \bibinfo {author} {\bibfnamefont
  {Y.}~\bibnamefont {Paltiel}}, \bibinfo {author} {\bibfnamefont
  {W.}~\bibnamefont {Petuskey}}, \bibinfo {author} {\bibfnamefont {J.~C.}\
  \bibnamefont {Ribeiro-Silva}}, \bibinfo {author} {\bibfnamefont
  {D.}~\bibnamefont {Stemer}}, \bibinfo {author} {\bibfnamefont
  {A.}~\bibnamefont {Valdes-Curiel}}, \bibinfo {author} {\bibfnamefont
  {S.}~\bibnamefont {Varela}}, \bibinfo {author} {\bibfnamefont
  {D.}~\bibnamefont {Waldeck}}, \bibinfo {author} {\bibfnamefont {P.~S.}\
  \bibnamefont {Weiss}}, \bibinfo {author} {\bibfnamefont {H.}~\bibnamefont
  {Zacharias}},\ and\ \bibinfo {author} {\bibfnamefont {Q.~H.}\ \bibnamefont
  {Wang}},\ }\href@noop {} {\bibinfo {title} {A chirality-based quantum leap: A
  forward-looking review}} (\bibinfo {year} {2020}),\ \Eprint
  {https://arxiv.org/abs/2009.00136} {arXiv:2009.00136 [cond-mat.mes-hall]}
  \BibitemShut {NoStop}%
\bibitem [{\citenamefont {Naaman}\ and\ \citenamefont
  {Waldeck}(2015)}]{naaman_spintronics_2015}%
  \BibitemOpen
  \bibfield  {author} {\bibinfo {author} {\bibfnamefont {R.}~\bibnamefont
  {Naaman}}\ and\ \bibinfo {author} {\bibfnamefont {D.~H.}\ \bibnamefont
  {Waldeck}},\ }\href {https://doi.org/10.1146/annurev-physchem-040214-121554}
  {\bibfield  {journal} {\bibinfo  {journal} {Annual Review of Physical
  Chemistry}\ }\textbf {\bibinfo {volume} {66}},\ \bibinfo {pages} {263}
  (\bibinfo {year} {2015})}\BibitemShut {NoStop}%
\bibitem [{\citenamefont {G{\"o}hler}\ \emph {et~al.}(2011)\citenamefont
  {G{\"o}hler}, \citenamefont {Hamelbeck}, \citenamefont {Markus},
  \citenamefont {Kettner}, \citenamefont {Hanne}, \citenamefont {Vager},
  \citenamefont {Naaman},\ and\ \citenamefont {Zacharias}}]{Goehler894}%
  \BibitemOpen
  \bibfield  {author} {\bibinfo {author} {\bibfnamefont {B.}~\bibnamefont
  {G{\"o}hler}}, \bibinfo {author} {\bibfnamefont {V.}~\bibnamefont
  {Hamelbeck}}, \bibinfo {author} {\bibfnamefont {T.~Z.}\ \bibnamefont
  {Markus}}, \bibinfo {author} {\bibfnamefont {M.}~\bibnamefont {Kettner}},
  \bibinfo {author} {\bibfnamefont {G.~F.}\ \bibnamefont {Hanne}}, \bibinfo
  {author} {\bibfnamefont {Z.}~\bibnamefont {Vager}}, \bibinfo {author}
  {\bibfnamefont {R.}~\bibnamefont {Naaman}},\ and\ \bibinfo {author}
  {\bibfnamefont {H.}~\bibnamefont {Zacharias}},\ }\href
  {https://doi.org/10.1126/science.1199339} {\bibfield  {journal} {\bibinfo
  {journal} {Science}\ }\textbf {\bibinfo {volume} {331}},\ \bibinfo {pages}
  {894} (\bibinfo {year} {2011})}\BibitemShut {NoStop}%
\bibitem [{\citenamefont {Kettner}\ \emph {et~al.}(2015)\citenamefont
  {Kettner}, \citenamefont {Göhler}, \citenamefont {Zacharias}, \citenamefont
  {Mishra}, \citenamefont {Kiran}, \citenamefont {Naaman}, \citenamefont
  {Fontanesi}, \citenamefont {Waldeck}, \citenamefont {Sęk}, \citenamefont
  {Pawłowski},\ and\ \citenamefont {Juhaniewicz}}]{kettner_spin_2015}%
  \BibitemOpen
  \bibfield  {author} {\bibinfo {author} {\bibfnamefont {M.}~\bibnamefont
  {Kettner}}, \bibinfo {author} {\bibfnamefont {B.}~\bibnamefont {Göhler}},
  \bibinfo {author} {\bibfnamefont {H.}~\bibnamefont {Zacharias}}, \bibinfo
  {author} {\bibfnamefont {D.}~\bibnamefont {Mishra}}, \bibinfo {author}
  {\bibfnamefont {V.}~\bibnamefont {Kiran}}, \bibinfo {author} {\bibfnamefont
  {R.}~\bibnamefont {Naaman}}, \bibinfo {author} {\bibfnamefont
  {C.}~\bibnamefont {Fontanesi}}, \bibinfo {author} {\bibfnamefont {D.~H.}\
  \bibnamefont {Waldeck}}, \bibinfo {author} {\bibfnamefont {S.}~\bibnamefont
  {Sęk}}, \bibinfo {author} {\bibfnamefont {J.}~\bibnamefont {Pawłowski}},\
  and\ \bibinfo {author} {\bibfnamefont {J.}~\bibnamefont {Juhaniewicz}},\
  }\href {https://doi.org/10.1021/jp509974z} {\bibfield  {journal} {\bibinfo
  {journal} {The Journal of Physical Chemistry C}\ }\textbf {\bibinfo {volume}
  {119}},\ \bibinfo {pages} {14542} (\bibinfo {year} {2015})}\BibitemShut
  {NoStop}%
\bibitem [{\citenamefont {Mishra}\ \emph {et~al.}(2013)\citenamefont {Mishra},
  \citenamefont {Markus}, \citenamefont {Naaman}, \citenamefont {Kettner},
  \citenamefont {Gohler}, \citenamefont {Zacharias}, \citenamefont {Friedman},
  \citenamefont {Sheves},\ and\ \citenamefont
  {Fontanesi}}]{mishra_spin-dependent_2013}%
  \BibitemOpen
  \bibfield  {author} {\bibinfo {author} {\bibfnamefont {D.}~\bibnamefont
  {Mishra}}, \bibinfo {author} {\bibfnamefont {T.~Z.}\ \bibnamefont {Markus}},
  \bibinfo {author} {\bibfnamefont {R.}~\bibnamefont {Naaman}}, \bibinfo
  {author} {\bibfnamefont {M.}~\bibnamefont {Kettner}}, \bibinfo {author}
  {\bibfnamefont {B.}~\bibnamefont {Gohler}}, \bibinfo {author} {\bibfnamefont
  {H.}~\bibnamefont {Zacharias}}, \bibinfo {author} {\bibfnamefont
  {N.}~\bibnamefont {Friedman}}, \bibinfo {author} {\bibfnamefont
  {M.}~\bibnamefont {Sheves}},\ and\ \bibinfo {author} {\bibfnamefont
  {C.}~\bibnamefont {Fontanesi}},\ }\href
  {https://doi.org/10.1073/pnas.1311493110} {\bibfield  {journal} {\bibinfo
  {journal} {Proceedings of the National Academy of Sciences}\ }\textbf
  {\bibinfo {volume} {110}},\ \bibinfo {pages} {14872} (\bibinfo {year}
  {2013})}\BibitemShut {NoStop}%
\bibitem [{\citenamefont {Carmeli}\ \emph {et~al.}(2014)\citenamefont
  {Carmeli}, \citenamefont {Kumar}, \citenamefont {Heifler}, \citenamefont
  {Carmeli},\ and\ \citenamefont {Naaman}}]{carmeli_spin_2014}%
  \BibitemOpen
  \bibfield  {author} {\bibinfo {author} {\bibfnamefont {I.}~\bibnamefont
  {Carmeli}}, \bibinfo {author} {\bibfnamefont {K.~S.}\ \bibnamefont {Kumar}},
  \bibinfo {author} {\bibfnamefont {O.}~\bibnamefont {Heifler}}, \bibinfo
  {author} {\bibfnamefont {C.}~\bibnamefont {Carmeli}},\ and\ \bibinfo {author}
  {\bibfnamefont {R.}~\bibnamefont {Naaman}},\ }\href
  {https://doi.org/10.1002/anie.201404382} {\bibfield  {journal} {\bibinfo
  {journal} {Angewandte Chemie International Edition}\ }\textbf {\bibinfo
  {volume} {53}},\ \bibinfo {pages} {8953} (\bibinfo {year}
  {2014})}\BibitemShut {NoStop}%
\bibitem [{\citenamefont {Niño}\ \emph {et~al.}(2014)\citenamefont {Niño},
  \citenamefont {Kowalik}, \citenamefont {Luque}, \citenamefont {Arvanitis},
  \citenamefont {Miranda},\ and\ \citenamefont
  {De~Miguel}}]{EnantiospecificSpinPol_2014}%
  \BibitemOpen
  \bibfield  {author} {\bibinfo {author} {\bibfnamefont {M.}~\bibnamefont
  {Niño}}, \bibinfo {author} {\bibfnamefont {I.}~\bibnamefont {Kowalik}},
  \bibinfo {author} {\bibfnamefont {F.}~\bibnamefont {Luque}}, \bibinfo
  {author} {\bibfnamefont {D.}~\bibnamefont {Arvanitis}}, \bibinfo {author}
  {\bibfnamefont {R.}~\bibnamefont {Miranda}},\ and\ \bibinfo {author}
  {\bibfnamefont {J.}~\bibnamefont {De~Miguel}},\ }\href
  {https://doi.org/10.1002/adma.201402810} {\bibfield  {journal} {\bibinfo
  {journal} {Advanced Materials}\ }\textbf {\bibinfo {volume} {26}} (\bibinfo
  {year} {2014})}\BibitemShut {NoStop}%
\bibitem [{\citenamefont {Kettner}\ \emph {et~al.}(2018)\citenamefont
  {Kettner}, \citenamefont {Maslyuk}, \citenamefont {Nürenberg}, \citenamefont
  {Seibel}, \citenamefont {Gutierrez}, \citenamefont {Cuniberti}, \citenamefont
  {Ernst},\ and\ \citenamefont {Zacharias}}]{kettner_2018}%
  \BibitemOpen
  \bibfield  {author} {\bibinfo {author} {\bibfnamefont {M.}~\bibnamefont
  {Kettner}}, \bibinfo {author} {\bibfnamefont {V.~V.}\ \bibnamefont
  {Maslyuk}}, \bibinfo {author} {\bibfnamefont {D.}~\bibnamefont {Nürenberg}},
  \bibinfo {author} {\bibfnamefont {J.}~\bibnamefont {Seibel}}, \bibinfo
  {author} {\bibfnamefont {R.}~\bibnamefont {Gutierrez}}, \bibinfo {author}
  {\bibfnamefont {G.}~\bibnamefont {Cuniberti}}, \bibinfo {author}
  {\bibfnamefont {K.-H.}\ \bibnamefont {Ernst}},\ and\ \bibinfo {author}
  {\bibfnamefont {H.}~\bibnamefont {Zacharias}},\ }\href
  {https://doi.org/10.1021/acs.jpclett.8b00208} {\bibfield  {journal} {\bibinfo
   {journal} {The Journal of Physical Chemistry Letters}\ }\textbf {\bibinfo
  {volume} {9}},\ \bibinfo {pages} {2025} (\bibinfo {year} {2018})}\BibitemShut
  {NoStop}%
\bibitem [{\citenamefont {Xie}\ \emph {et~al.}(2011)\citenamefont {Xie},
  \citenamefont {Markus}, \citenamefont {Cohen}, \citenamefont {Vager},
  \citenamefont {Gutierrez},\ and\ \citenamefont
  {Naaman}}]{dna_oligomers_2011}%
  \BibitemOpen
  \bibfield  {author} {\bibinfo {author} {\bibfnamefont {Z.}~\bibnamefont
  {Xie}}, \bibinfo {author} {\bibfnamefont {T.}~\bibnamefont {Markus}},
  \bibinfo {author} {\bibfnamefont {S.}~\bibnamefont {Cohen}}, \bibinfo
  {author} {\bibfnamefont {Z.}~\bibnamefont {Vager}}, \bibinfo {author}
  {\bibfnamefont {R.}~\bibnamefont {Gutierrez}},\ and\ \bibinfo {author}
  {\bibfnamefont {R.}~\bibnamefont {Naaman}},\ }\href
  {https://doi.org/10.1021/nl2021637} {\bibfield  {journal} {\bibinfo
  {journal} {Nano Letters}\ }\textbf {\bibinfo {volume} {11}},\ \bibinfo
  {pages} {4652} (\bibinfo {year} {2011})}\BibitemShut {NoStop}%
\bibitem [{\citenamefont {Bostick}\ \emph {et~al.}(2018)\citenamefont
  {Bostick}, \citenamefont {Mukhopadhyay}, \citenamefont {Pecht}, \citenamefont
  {Sheves}, \citenamefont {Cahen},\ and\ \citenamefont
  {Lederman}}]{bostick_2018}%
  \BibitemOpen
  \bibfield  {author} {\bibinfo {author} {\bibfnamefont {C.~D.}\ \bibnamefont
  {Bostick}}, \bibinfo {author} {\bibfnamefont {S.}~\bibnamefont
  {Mukhopadhyay}}, \bibinfo {author} {\bibfnamefont {I.}~\bibnamefont {Pecht}},
  \bibinfo {author} {\bibfnamefont {M.}~\bibnamefont {Sheves}}, \bibinfo
  {author} {\bibfnamefont {D.}~\bibnamefont {Cahen}},\ and\ \bibinfo {author}
  {\bibfnamefont {D.}~\bibnamefont {Lederman}},\ }\href
  {https://doi.org/10.1088/1361-6633/aa85f2} {\bibfield  {journal} {\bibinfo
  {journal} {Reports on Progress in Physics}\ }\textbf {\bibinfo {volume}
  {81}},\ \bibinfo {pages} {026601} (\bibinfo {year} {2018})}\BibitemShut
  {NoStop}%
\bibitem [{\citenamefont {Dor}\ \emph {et~al.}(2013)\citenamefont {Dor},
  \citenamefont {Yochelis}, \citenamefont {Mathew}, \citenamefont {Naaman},\
  and\ \citenamefont {Paltiel}}]{dor_chiral-based_2013}%
  \BibitemOpen
  \bibfield  {author} {\bibinfo {author} {\bibfnamefont {O.~B.}\ \bibnamefont
  {Dor}}, \bibinfo {author} {\bibfnamefont {S.}~\bibnamefont {Yochelis}},
  \bibinfo {author} {\bibfnamefont {S.~P.}\ \bibnamefont {Mathew}}, \bibinfo
  {author} {\bibfnamefont {R.}~\bibnamefont {Naaman}},\ and\ \bibinfo {author}
  {\bibfnamefont {Y.}~\bibnamefont {Paltiel}},\ }\href
  {https://doi.org/10.1038/ncomms3256} {\bibfield  {journal} {\bibinfo
  {journal} {Nature Communications}\ }\textbf {\bibinfo {volume} {4}},\
  \bibinfo {pages} {2256} (\bibinfo {year} {2013})}\BibitemShut {NoStop}%
\bibitem [{\citenamefont {Michaeli}\ \emph {et~al.}(2017)\citenamefont
  {Michaeli}, \citenamefont {Varade}, \citenamefont {Naaman},\ and\
  \citenamefont {Waldeck}}]{Michaeli_2017}%
  \BibitemOpen
  \bibfield  {author} {\bibinfo {author} {\bibfnamefont {K.}~\bibnamefont
  {Michaeli}}, \bibinfo {author} {\bibfnamefont {V.}~\bibnamefont {Varade}},
  \bibinfo {author} {\bibfnamefont {R.}~\bibnamefont {Naaman}},\ and\ \bibinfo
  {author} {\bibfnamefont {D.~H.}\ \bibnamefont {Waldeck}},\ }\href
  {https://doi.org/10.1088/1361-648x/aa54a4} {\bibfield  {journal} {\bibinfo
  {journal} {Journal of Physics: Condensed Matter}\ }\textbf {\bibinfo {volume}
  {29}},\ \bibinfo {pages} {103002} (\bibinfo {year} {2017})}\BibitemShut
  {NoStop}%
\bibitem [{\citenamefont {Mondal}\ \emph {et~al.}(2016)\citenamefont {Mondal},
  \citenamefont {Fontanesi}, \citenamefont {Waldeck},\ and\ \citenamefont
  {Naaman}}]{mondal_spin-dependent_2016}%
  \BibitemOpen
  \bibfield  {author} {\bibinfo {author} {\bibfnamefont {P.~C.}\ \bibnamefont
  {Mondal}}, \bibinfo {author} {\bibfnamefont {C.}~\bibnamefont {Fontanesi}},
  \bibinfo {author} {\bibfnamefont {D.~H.}\ \bibnamefont {Waldeck}},\ and\
  \bibinfo {author} {\bibfnamefont {R.}~\bibnamefont {Naaman}},\ }\href
  {https://doi.org/10.1021/acs.accounts.6b00446} {\bibfield  {journal}
  {\bibinfo  {journal} {Accounts of Chemical Research}\ }\textbf {\bibinfo
  {volume} {49}},\ \bibinfo {pages} {2560} (\bibinfo {year}
  {2016})}\BibitemShut {NoStop}%
\bibitem [{\citenamefont {Yang}\ \emph {et~al.}(2020)\citenamefont {Yang},
  \citenamefont {van~der Wal},\ and\ \citenamefont {van Wees}}]{yang_2020}%
  \BibitemOpen
  \bibfield  {author} {\bibinfo {author} {\bibfnamefont {X.}~\bibnamefont
  {Yang}}, \bibinfo {author} {\bibfnamefont {C.~H.}\ \bibnamefont {van~der
  Wal}},\ and\ \bibinfo {author} {\bibfnamefont {B.~J.}\ \bibnamefont {van
  Wees}},\ }\href {https://doi.org/10.1021/acs.nanolett.0c02417} {\bibfield
  {journal} {\bibinfo  {journal} {Nano Letters}\ }\textbf {\bibinfo {volume}
  {20}},\ \bibinfo {pages} {6148} (\bibinfo {year} {2020})}\BibitemShut
  {NoStop}%
\bibitem [{\citenamefont {Chiesa}\ \emph {et~al.}(2021)\citenamefont {Chiesa},
  \citenamefont {Chizzini}, \citenamefont {Garlatti}, \citenamefont
  {Salvadori}, \citenamefont {Tacchino}, \citenamefont {Santini}, \citenamefont
  {Tavernelli}, \citenamefont {Bittl}, \citenamefont {Chiesa}, \citenamefont
  {Sessoli},\ and\ \citenamefont {Carretta}}]{Chiesa2021}%
  \BibitemOpen
  \bibfield  {author} {\bibinfo {author} {\bibfnamefont {A.}~\bibnamefont
  {Chiesa}}, \bibinfo {author} {\bibfnamefont {M.}~\bibnamefont {Chizzini}},
  \bibinfo {author} {\bibfnamefont {E.}~\bibnamefont {Garlatti}}, \bibinfo
  {author} {\bibfnamefont {E.}~\bibnamefont {Salvadori}}, \bibinfo {author}
  {\bibfnamefont {F.}~\bibnamefont {Tacchino}}, \bibinfo {author}
  {\bibfnamefont {P.}~\bibnamefont {Santini}}, \bibinfo {author} {\bibfnamefont
  {I.}~\bibnamefont {Tavernelli}}, \bibinfo {author} {\bibfnamefont
  {R.}~\bibnamefont {Bittl}}, \bibinfo {author} {\bibfnamefont
  {M.}~\bibnamefont {Chiesa}}, \bibinfo {author} {\bibfnamefont
  {R.}~\bibnamefont {Sessoli}},\ and\ \bibinfo {author} {\bibfnamefont
  {S.}~\bibnamefont {Carretta}},\ }\href
  {https://doi.org/10.1021/acs.jpclett.1c01447} {\bibfield  {journal} {\bibinfo
   {journal} {The Journal of Physical Chemistry Letters}\ }\textbf {\bibinfo
  {volume} {12}},\ \bibinfo {pages} {6341} (\bibinfo {year}
  {2021})}\BibitemShut {NoStop}%
\bibitem [{\citenamefont {Kumar}\ \emph {et~al.}(2017)\citenamefont {Kumar},
  \citenamefont {Capua}, \citenamefont {Kesharwani}, \citenamefont {Martin},
  \citenamefont {Sitbon}, \citenamefont {Waldeck},\ and\ \citenamefont
  {Naaman}}]{kumar_chirality-induced_2017}%
  \BibitemOpen
  \bibfield  {author} {\bibinfo {author} {\bibfnamefont {A.}~\bibnamefont
  {Kumar}}, \bibinfo {author} {\bibfnamefont {E.}~\bibnamefont {Capua}},
  \bibinfo {author} {\bibfnamefont {M.~K.}\ \bibnamefont {Kesharwani}},
  \bibinfo {author} {\bibfnamefont {J.~M.~L.}\ \bibnamefont {Martin}}, \bibinfo
  {author} {\bibfnamefont {E.}~\bibnamefont {Sitbon}}, \bibinfo {author}
  {\bibfnamefont {D.~H.}\ \bibnamefont {Waldeck}},\ and\ \bibinfo {author}
  {\bibfnamefont {R.}~\bibnamefont {Naaman}},\ }\href
  {https://doi.org/10.1073/pnas.1611467114} {\bibfield  {journal} {\bibinfo
  {journal} {Proceedings of the National Academy of Sciences}\ }\textbf
  {\bibinfo {volume} {114}},\ \bibinfo {pages} {2474} (\bibinfo {year}
  {2017})}\BibitemShut {NoStop}%
\bibitem [{\citenamefont {Banerjee-Ghosh}\ \emph {et~al.}(2018)\citenamefont
  {Banerjee-Ghosh}, \citenamefont {Dor}, \citenamefont {Tassinari},
  \citenamefont {Capua}, \citenamefont {Yochelis}, \citenamefont {Capua},
  \citenamefont {Yang}, \citenamefont {Parkin}, \citenamefont {Sarkar},
  \citenamefont {Kronik}, \citenamefont {Baczewski}, \citenamefont {Naaman},\
  and\ \citenamefont {Paltiel}}]{banerhee-ghosh_2018}%
  \BibitemOpen
  \bibfield  {author} {\bibinfo {author} {\bibfnamefont {K.}~\bibnamefont
  {Banerjee-Ghosh}}, \bibinfo {author} {\bibfnamefont {O.~B.}\ \bibnamefont
  {Dor}}, \bibinfo {author} {\bibfnamefont {F.}~\bibnamefont {Tassinari}},
  \bibinfo {author} {\bibfnamefont {E.}~\bibnamefont {Capua}}, \bibinfo
  {author} {\bibfnamefont {S.}~\bibnamefont {Yochelis}}, \bibinfo {author}
  {\bibfnamefont {A.}~\bibnamefont {Capua}}, \bibinfo {author} {\bibfnamefont
  {S.-H.}\ \bibnamefont {Yang}}, \bibinfo {author} {\bibfnamefont {S.~S.~P.}\
  \bibnamefont {Parkin}}, \bibinfo {author} {\bibfnamefont {S.}~\bibnamefont
  {Sarkar}}, \bibinfo {author} {\bibfnamefont {L.}~\bibnamefont {Kronik}},
  \bibinfo {author} {\bibfnamefont {L.~T.}\ \bibnamefont {Baczewski}}, \bibinfo
  {author} {\bibfnamefont {R.}~\bibnamefont {Naaman}},\ and\ \bibinfo {author}
  {\bibfnamefont {Y.}~\bibnamefont {Paltiel}},\ }\href
  {https://doi.org/10.1126/science.aar4265} {\bibfield  {journal} {\bibinfo
  {journal} {Science}\ }\textbf {\bibinfo {volume} {360}},\ \bibinfo {pages}
  {1331} (\bibinfo {year} {2018})}\BibitemShut {NoStop}%
\bibitem [{\citenamefont {Metzger}\ \emph {et~al.}(2020)\citenamefont
  {Metzger}, \citenamefont {Mishra}, \citenamefont {Bloom}, \citenamefont
  {Goren}, \citenamefont {Neubauer}, \citenamefont {Shmul}, \citenamefont
  {Wei}, \citenamefont {Yochelis}, \citenamefont {Tassinari}, \citenamefont
  {Fontanesi}, \citenamefont {Waldeck}, \citenamefont {Paltiel},\ and\
  \citenamefont {Naaman}}]{metzger_electron_2020}%
  \BibitemOpen
  \bibfield  {author} {\bibinfo {author} {\bibfnamefont {T.~S.}\ \bibnamefont
  {Metzger}}, \bibinfo {author} {\bibfnamefont {S.}~\bibnamefont {Mishra}},
  \bibinfo {author} {\bibfnamefont {B.~P.}\ \bibnamefont {Bloom}}, \bibinfo
  {author} {\bibfnamefont {N.}~\bibnamefont {Goren}}, \bibinfo {author}
  {\bibfnamefont {A.}~\bibnamefont {Neubauer}}, \bibinfo {author}
  {\bibfnamefont {G.}~\bibnamefont {Shmul}}, \bibinfo {author} {\bibfnamefont
  {J.}~\bibnamefont {Wei}}, \bibinfo {author} {\bibfnamefont {S.}~\bibnamefont
  {Yochelis}}, \bibinfo {author} {\bibfnamefont {F.}~\bibnamefont {Tassinari}},
  \bibinfo {author} {\bibfnamefont {C.}~\bibnamefont {Fontanesi}}, \bibinfo
  {author} {\bibfnamefont {D.~H.}\ \bibnamefont {Waldeck}}, \bibinfo {author}
  {\bibfnamefont {Y.}~\bibnamefont {Paltiel}},\ and\ \bibinfo {author}
  {\bibfnamefont {R.}~\bibnamefont {Naaman}},\ }\href
  {https://doi.org/10.1002/anie.201911400} {\bibfield  {journal} {\bibinfo
  {journal} {Angewandte Chemie International Edition}\ }\textbf {\bibinfo
  {volume} {59}},\ \bibinfo {pages} {1653} (\bibinfo {year}
  {2020})}\BibitemShut {NoStop}%
\bibitem [{\citenamefont {Dianat}\ \emph {et~al.}(2020)\citenamefont {Dianat},
  \citenamefont {Gutierrez}, \citenamefont {Alpern}, \citenamefont {Mujica},
  \citenamefont {Ziv}, \citenamefont {Yochelis}, \citenamefont {Millo},
  \citenamefont {Paltiel},\ and\ \citenamefont {Cuniberti}}]{dianat_2020}%
  \BibitemOpen
  \bibfield  {author} {\bibinfo {author} {\bibfnamefont {A.}~\bibnamefont
  {Dianat}}, \bibinfo {author} {\bibfnamefont {R.}~\bibnamefont {Gutierrez}},
  \bibinfo {author} {\bibfnamefont {H.}~\bibnamefont {Alpern}}, \bibinfo
  {author} {\bibfnamefont {V.}~\bibnamefont {Mujica}}, \bibinfo {author}
  {\bibfnamefont {A.}~\bibnamefont {Ziv}}, \bibinfo {author} {\bibfnamefont
  {S.}~\bibnamefont {Yochelis}}, \bibinfo {author} {\bibfnamefont
  {O.}~\bibnamefont {Millo}}, \bibinfo {author} {\bibfnamefont
  {Y.}~\bibnamefont {Paltiel}},\ and\ \bibinfo {author} {\bibfnamefont
  {G.}~\bibnamefont {Cuniberti}},\ }\href
  {https://doi.org/10.1021/acs.nanolett.0c02216} {\bibfield  {journal}
  {\bibinfo  {journal} {Nano Letters}\ }\textbf {\bibinfo {volume} {20}},\
  \bibinfo {pages} {7077} (\bibinfo {year} {2020})}\BibitemShut {NoStop}%
\bibitem [{\citenamefont {Kapon}\ \emph {et~al.}(2021)\citenamefont {Kapon},
  \citenamefont {Saha}, \citenamefont {Duanis-Assaf}, \citenamefont {Stuyver},
  \citenamefont {Ziv}, \citenamefont {Metzger}, \citenamefont {Yochelis},
  \citenamefont {Shaik}, \citenamefont {Naaman}, \citenamefont {Reches},\ and\
  \citenamefont {Paltiel}}]{kapon_evidence_2021}%
  \BibitemOpen
  \bibfield  {author} {\bibinfo {author} {\bibfnamefont {Y.}~\bibnamefont
  {Kapon}}, \bibinfo {author} {\bibfnamefont {A.}~\bibnamefont {Saha}},
  \bibinfo {author} {\bibfnamefont {T.}~\bibnamefont {Duanis-Assaf}}, \bibinfo
  {author} {\bibfnamefont {T.}~\bibnamefont {Stuyver}}, \bibinfo {author}
  {\bibfnamefont {A.}~\bibnamefont {Ziv}}, \bibinfo {author} {\bibfnamefont
  {T.}~\bibnamefont {Metzger}}, \bibinfo {author} {\bibfnamefont
  {S.}~\bibnamefont {Yochelis}}, \bibinfo {author} {\bibfnamefont
  {S.}~\bibnamefont {Shaik}}, \bibinfo {author} {\bibfnamefont
  {R.}~\bibnamefont {Naaman}}, \bibinfo {author} {\bibfnamefont
  {M.}~\bibnamefont {Reches}},\ and\ \bibinfo {author} {\bibfnamefont
  {Y.}~\bibnamefont {Paltiel}},\ }\href
  {https://doi.org/https://doi.org/10.1016/j.chempr.2021.08.002} {\bibfield
  {journal} {\bibinfo  {journal} {Chem}\ }\textbf {\bibinfo {volume} {7}},\
  \bibinfo {pages} {2787} (\bibinfo {year} {2021})}\BibitemShut {NoStop}%
\bibitem [{\citenamefont {Evers}\ \emph {et~al.}(2021)\citenamefont {Evers},
  \citenamefont {Aharony}, \citenamefont {Bar-Gill}, \citenamefont
  {Entin-Wohlman}, \citenamefont {Hedegård}, \citenamefont {Hod},
  \citenamefont {Jelinek}, \citenamefont {Kamieniarz}, \citenamefont
  {Lemeshko}, \citenamefont {Michaeli}, \citenamefont {Mujica}, \citenamefont
  {Naaman}, \citenamefont {Paltiel}, \citenamefont {Refaely-Abramson},
  \citenamefont {Tal}, \citenamefont {Thijssen}, \citenamefont {Thoss},
  \citenamefont {van Ruitenbeek}, \citenamefont {Venkataraman}, \citenamefont
  {Waldeck}, \citenamefont {Yan},\ and\ \citenamefont {Kronik}}]{Overview2021}%
  \BibitemOpen
  \bibfield  {author} {\bibinfo {author} {\bibfnamefont {F.}~\bibnamefont
  {Evers}}, \bibinfo {author} {\bibfnamefont {A.}~\bibnamefont {Aharony}},
  \bibinfo {author} {\bibfnamefont {N.}~\bibnamefont {Bar-Gill}}, \bibinfo
  {author} {\bibfnamefont {O.}~\bibnamefont {Entin-Wohlman}}, \bibinfo {author}
  {\bibfnamefont {P.}~\bibnamefont {Hedegård}}, \bibinfo {author}
  {\bibfnamefont {O.}~\bibnamefont {Hod}}, \bibinfo {author} {\bibfnamefont
  {P.}~\bibnamefont {Jelinek}}, \bibinfo {author} {\bibfnamefont
  {G.}~\bibnamefont {Kamieniarz}}, \bibinfo {author} {\bibfnamefont
  {M.}~\bibnamefont {Lemeshko}}, \bibinfo {author} {\bibfnamefont
  {K.}~\bibnamefont {Michaeli}}, \bibinfo {author} {\bibfnamefont
  {V.}~\bibnamefont {Mujica}}, \bibinfo {author} {\bibfnamefont
  {R.}~\bibnamefont {Naaman}}, \bibinfo {author} {\bibfnamefont
  {Y.}~\bibnamefont {Paltiel}}, \bibinfo {author} {\bibfnamefont
  {S.}~\bibnamefont {Refaely-Abramson}}, \bibinfo {author} {\bibfnamefont
  {O.}~\bibnamefont {Tal}}, \bibinfo {author} {\bibfnamefont {J.}~\bibnamefont
  {Thijssen}}, \bibinfo {author} {\bibfnamefont {M.}~\bibnamefont {Thoss}},
  \bibinfo {author} {\bibfnamefont {J.~M.}\ \bibnamefont {van Ruitenbeek}},
  \bibinfo {author} {\bibfnamefont {L.}~\bibnamefont {Venkataraman}}, \bibinfo
  {author} {\bibfnamefont {D.~H.}\ \bibnamefont {Waldeck}}, \bibinfo {author}
  {\bibfnamefont {B.}~\bibnamefont {Yan}},\ and\ \bibinfo {author}
  {\bibfnamefont {L.}~\bibnamefont {Kronik}},\ }\href@noop {} {\bibinfo {title}
  {Theory of chirality induced spin selectivity: Progress and challenges}}
  (\bibinfo {year} {2021}),\ \Eprint {https://arxiv.org/abs/2108.09998}
  {arXiv:2108.09998 [cond-mat.mtrl-sci]} \BibitemShut {NoStop}%
\bibitem [{\citenamefont {Naaman}\ and\ \citenamefont
  {Waldeck}(2012)}]{naaman_chiral-induced_2012}%
  \BibitemOpen
  \bibfield  {author} {\bibinfo {author} {\bibfnamefont {R.}~\bibnamefont
  {Naaman}}\ and\ \bibinfo {author} {\bibfnamefont {D.~H.}\ \bibnamefont
  {Waldeck}},\ }\href {https://doi.org/10.1021/jz300793y} {\bibfield  {journal}
  {\bibinfo  {journal} {The Journal of Physical Chemistry Letters}\ }\textbf
  {\bibinfo {volume} {3}},\ \bibinfo {pages} {2178} (\bibinfo {year}
  {2012})}\BibitemShut {NoStop}%
\bibitem [{\citenamefont {Gutierrez}\ \emph {et~al.}(2012)\citenamefont
  {Gutierrez}, \citenamefont {D\'{\i}az}, \citenamefont {Naaman},\ and\
  \citenamefont {Cuniberti}}]{Cuniberti2012}%
  \BibitemOpen
  \bibfield  {author} {\bibinfo {author} {\bibfnamefont {R.}~\bibnamefont
  {Gutierrez}}, \bibinfo {author} {\bibfnamefont {E.}~\bibnamefont
  {D\'{\i}az}}, \bibinfo {author} {\bibfnamefont {R.}~\bibnamefont {Naaman}},\
  and\ \bibinfo {author} {\bibfnamefont {G.}~\bibnamefont {Cuniberti}},\ }\href
  {https://doi.org/10.1103/PhysRevB.85.081404} {\bibfield  {journal} {\bibinfo
  {journal} {Phys. Rev. B}\ }\textbf {\bibinfo {volume} {85}},\ \bibinfo
  {pages} {081404} (\bibinfo {year} {2012})}\BibitemShut {NoStop}%
\bibitem [{\citenamefont {Guo}\ and\ \citenamefont
  {Sun}(2014)}]{guo_spin-dependent_2014}%
  \BibitemOpen
  \bibfield  {author} {\bibinfo {author} {\bibfnamefont {A.-M.}\ \bibnamefont
  {Guo}}\ and\ \bibinfo {author} {\bibfnamefont {Q.-F.}\ \bibnamefont {Sun}},\
  }\href {https://doi.org/10.1073/pnas.1407716111} {\bibfield  {journal}
  {\bibinfo  {journal} {Proceedings of the National Academy of Sciences}\
  }\textbf {\bibinfo {volume} {111}},\ \bibinfo {pages} {11658} (\bibinfo
  {year} {2014})}\BibitemShut {NoStop}%
\bibitem [{\citenamefont {Gutierrez}\ \emph {et~al.}(2013)\citenamefont
  {Gutierrez}, \citenamefont {Díaz}, \citenamefont {Gaul}, \citenamefont
  {Brumme}, \citenamefont {Domínguez-Adame},\ and\ \citenamefont
  {Cuniberti}}]{GutierrezCunibertiEffective}%
  \BibitemOpen
  \bibfield  {author} {\bibinfo {author} {\bibfnamefont {R.}~\bibnamefont
  {Gutierrez}}, \bibinfo {author} {\bibfnamefont {E.}~\bibnamefont {Díaz}},
  \bibinfo {author} {\bibfnamefont {C.}~\bibnamefont {Gaul}}, \bibinfo {author}
  {\bibfnamefont {T.}~\bibnamefont {Brumme}}, \bibinfo {author} {\bibfnamefont
  {F.}~\bibnamefont {Domínguez-Adame}},\ and\ \bibinfo {author} {\bibfnamefont
  {G.}~\bibnamefont {Cuniberti}},\ }\href {https://doi.org/10.1021/jp401705x}
  {\bibfield  {journal} {\bibinfo  {journal} {The Journal of Physical Chemistry
  C}\ }\textbf {\bibinfo {volume} {117}},\ \bibinfo {pages} {22276} (\bibinfo
  {year} {2013})}\BibitemShut {NoStop}%
\bibitem [{\citenamefont {Naaman}\ \emph {et~al.}(2019)\citenamefont {Naaman},
  \citenamefont {Paltiel},\ and\ \citenamefont {Waldeck}}]{naaman_chiral_2019}%
  \BibitemOpen
  \bibfield  {author} {\bibinfo {author} {\bibfnamefont {R.}~\bibnamefont
  {Naaman}}, \bibinfo {author} {\bibfnamefont {Y.}~\bibnamefont {Paltiel}},\
  and\ \bibinfo {author} {\bibfnamefont {D.~H.}\ \bibnamefont {Waldeck}},\
  }\href {https://doi.org/10.1038/s41570-019-0087-1} {\bibfield  {journal}
  {\bibinfo  {journal} {Nature Reviews Chemistry}\ }\textbf {\bibinfo {volume}
  {3}},\ \bibinfo {pages} {250} (\bibinfo {year} {2019})}\BibitemShut {NoStop}%
\bibitem [{\citenamefont {Michaeli}\ and\ \citenamefont
  {Naaman}(2019)}]{Naaman_Michaeli_2019}%
  \BibitemOpen
  \bibfield  {author} {\bibinfo {author} {\bibfnamefont {K.}~\bibnamefont
  {Michaeli}}\ and\ \bibinfo {author} {\bibfnamefont {R.}~\bibnamefont
  {Naaman}},\ }\href {https://doi.org/10.1021/acs.jpcc.9b05020} {\bibfield
  {journal} {\bibinfo  {journal} {The Journal of Physical Chemistry C}\
  }\textbf {\bibinfo {volume} {123}},\ \bibinfo {pages} {17043} (\bibinfo
  {year} {2019})}\BibitemShut {NoStop}%
\bibitem [{\citenamefont {Ghazaryan}\ \emph {et~al.}(2020)\citenamefont
  {Ghazaryan}, \citenamefont {Paltiel},\ and\ \citenamefont
  {Lemeshko}}]{GhazaryanAnalyticModel2020}%
  \BibitemOpen
  \bibfield  {author} {\bibinfo {author} {\bibfnamefont {A.}~\bibnamefont
  {Ghazaryan}}, \bibinfo {author} {\bibfnamefont {Y.}~\bibnamefont {Paltiel}},\
  and\ \bibinfo {author} {\bibfnamefont {M.}~\bibnamefont {Lemeshko}},\ }\href
  {https://doi.org/10.1021/acs.jpcc.0c02584} {\bibfield  {journal} {\bibinfo
  {journal} {The Journal of Physical Chemistry C}\ }\textbf {\bibinfo {volume}
  {124}},\ \bibinfo {pages} {11716} (\bibinfo {year} {2020})}\BibitemShut
  {NoStop}%
\bibitem [{\citenamefont {Ortix}(2015)}]{ortix_quantum_2015}%
  \BibitemOpen
  \bibfield  {author} {\bibinfo {author} {\bibfnamefont {C.}~\bibnamefont
  {Ortix}},\ }\href {https://doi.org/10.1103/PhysRevB.91.245412} {\bibfield
  {journal} {\bibinfo  {journal} {Physical Review B}\ }\textbf {\bibinfo
  {volume} {91}},\ \bibinfo {pages} {245412} (\bibinfo {year}
  {2015})}\BibitemShut {NoStop}%
\bibitem [{\citenamefont {Ladik}(1974)}]{EffectiveMass1}%
  \BibitemOpen
  \bibfield  {author} {\bibinfo {author} {\bibfnamefont {J.}~\bibnamefont
  {Ladik}},\ }\href {https://doi.org/10.1002/qua.560080711} {\bibfield
  {journal} {\bibinfo  {journal} {International Journal of Quantum Chemistry}\
  }\textbf {\bibinfo {volume} {8}},\ \bibinfo {pages} {65} (\bibinfo {year}
  {1974})}\BibitemShut {NoStop}%
\bibitem [{\citenamefont {Maia}\ \emph {et~al.}(2011)\citenamefont {Maia},
  \citenamefont {Freire}, \citenamefont {Caetano}, \citenamefont {Azevedo},
  \citenamefont {Sales},\ and\ \citenamefont {Albuquerque}}]{EffectiveMass2}%
  \BibitemOpen
  \bibfield  {author} {\bibinfo {author} {\bibfnamefont {F.~F.}\ \bibnamefont
  {Maia}}, \bibinfo {author} {\bibfnamefont {V.~N.}\ \bibnamefont {Freire}},
  \bibinfo {author} {\bibfnamefont {E.~W.~S.}\ \bibnamefont {Caetano}},
  \bibinfo {author} {\bibfnamefont {D.~L.}\ \bibnamefont {Azevedo}}, \bibinfo
  {author} {\bibfnamefont {F.~A.~M.}\ \bibnamefont {Sales}},\ and\ \bibinfo
  {author} {\bibfnamefont {E.~L.}\ \bibnamefont {Albuquerque}},\ }\href
  {https://doi.org/10.1063/1.3584680} {\bibfield  {journal} {\bibinfo
  {journal} {The Journal of Chemical Physics}\ }\textbf {\bibinfo {volume}
  {134}},\ \bibinfo {pages} {175101} (\bibinfo {year} {2011})}\BibitemShut
  {NoStop}%
\bibitem [{\citenamefont {Shankar}(1994)}]{Shankar}%
  \BibitemOpen
  \bibfield  {author} {\bibinfo {author} {\bibfnamefont {R.}~\bibnamefont
  {Shankar}},\ }\href@noop {} {\emph {\bibinfo {title} {Principles of Quantum
  Mechanics}}},\ \bibinfo {edition} {2nd}\ ed.\ (\bibinfo  {publisher} {Plenum
  Press},\ \bibinfo {address} {New York},\ \bibinfo {year} {1994})\
  Chap.~\bibinfo {chapter} {5}, pp.\ \bibinfo {pages} {167--175}\BibitemShut
  {NoStop}%
\bibitem [{\citenamefont {Entin-Wohlman}\ \emph {et~al.}(2021)\citenamefont
  {Entin-Wohlman}, \citenamefont {Aharony},\ and\ \citenamefont
  {Utsumi}}]{UtsumiComment}%
  \BibitemOpen
  \bibfield  {author} {\bibinfo {author} {\bibfnamefont {O.}~\bibnamefont
  {Entin-Wohlman}}, \bibinfo {author} {\bibfnamefont {A.}~\bibnamefont
  {Aharony}},\ and\ \bibinfo {author} {\bibfnamefont {Y.}~\bibnamefont
  {Utsumi}},\ }\href {https://doi.org/10.1103/PhysRevB.103.077401} {\bibfield
  {journal} {\bibinfo  {journal} {Phys. Rev. B}\ }\textbf {\bibinfo {volume}
  {103}},\ \bibinfo {pages} {077401} (\bibinfo {year} {2021})}\BibitemShut
  {NoStop}%
\bibitem [{\citenamefont {Guo}\ and\ \citenamefont {Sun}(2012)}]{Guo_2012}%
  \BibitemOpen
  \bibfield  {author} {\bibinfo {author} {\bibfnamefont {A.-M.}\ \bibnamefont
  {Guo}}\ and\ \bibinfo {author} {\bibfnamefont {Q.-F.}\ \bibnamefont {Sun}},\
  }\href {https://doi.org/10.1103/PhysRevLett.108.218102} {\bibfield  {journal}
  {\bibinfo  {journal} {Phys. Rev. Lett.}\ }\textbf {\bibinfo {volume} {108}},\
  \bibinfo {pages} {218102} (\bibinfo {year} {2012})}\BibitemShut {NoStop}%
\bibitem [{\citenamefont {Sierra}\ \emph {et~al.}(2019)\citenamefont {Sierra},
  \citenamefont {Sánchez}, \citenamefont {Gutierrez}, \citenamefont
  {Cuniberti}, \citenamefont {Domínguez-Adame},\ and\ \citenamefont
  {Díaz}}]{sierra_spin-polarized_2019}%
  \BibitemOpen
  \bibfield  {author} {\bibinfo {author} {\bibfnamefont {M.~A.}\ \bibnamefont
  {Sierra}}, \bibinfo {author} {\bibfnamefont {D.}~\bibnamefont {Sánchez}},
  \bibinfo {author} {\bibfnamefont {R.}~\bibnamefont {Gutierrez}}, \bibinfo
  {author} {\bibfnamefont {G.}~\bibnamefont {Cuniberti}}, \bibinfo {author}
  {\bibfnamefont {F.}~\bibnamefont {Domínguez-Adame}},\ and\ \bibinfo {author}
  {\bibfnamefont {E.}~\bibnamefont {Díaz}},\ }\href
  {https://doi.org/10.3390/biom10010049} {\bibfield  {journal} {\bibinfo
  {journal} {Biomolecules}\ }\textbf {\bibinfo {volume} {10}},\ \bibinfo
  {pages} {49} (\bibinfo {year} {2019})}\BibitemShut {NoStop}%
\bibitem [{\citenamefont {Utsumi}\ \emph {et~al.}(2020)\citenamefont {Utsumi},
  \citenamefont {Entin-Wohlman},\ and\ \citenamefont {Aharony}}]{Utsumi2020}%
  \BibitemOpen
  \bibfield  {author} {\bibinfo {author} {\bibfnamefont {Y.}~\bibnamefont
  {Utsumi}}, \bibinfo {author} {\bibfnamefont {O.}~\bibnamefont
  {Entin-Wohlman}},\ and\ \bibinfo {author} {\bibfnamefont {A.}~\bibnamefont
  {Aharony}},\ }\href {https://doi.org/10.1103/PhysRevB.102.035445} {\bibfield
  {journal} {\bibinfo  {journal} {Phys. Rev. B}\ }\textbf {\bibinfo {volume}
  {102}},\ \bibinfo {pages} {035445} (\bibinfo {year} {2020})}\BibitemShut
  {NoStop}%
\bibitem [{\citenamefont {Guo}\ \emph {et~al.}(2014)\citenamefont {Guo},
  \citenamefont {D\'{\i}az}, \citenamefont {Gaul}, \citenamefont {Gutierrez},
  \citenamefont {Dom\'{\i}nguez-Adame}, \citenamefont {Cuniberti},\ and\
  \citenamefont {Sun}}]{GuoContactEffects2014}%
  \BibitemOpen
  \bibfield  {author} {\bibinfo {author} {\bibfnamefont {A.-M.}\ \bibnamefont
  {Guo}}, \bibinfo {author} {\bibfnamefont {E.}~\bibnamefont {D\'{\i}az}},
  \bibinfo {author} {\bibfnamefont {C.}~\bibnamefont {Gaul}}, \bibinfo {author}
  {\bibfnamefont {R.}~\bibnamefont {Gutierrez}}, \bibinfo {author}
  {\bibfnamefont {F.}~\bibnamefont {Dom\'{\i}nguez-Adame}}, \bibinfo {author}
  {\bibfnamefont {G.}~\bibnamefont {Cuniberti}},\ and\ \bibinfo {author}
  {\bibfnamefont {Q.-f.}\ \bibnamefont {Sun}},\ }\href
  {https://doi.org/10.1103/PhysRevB.89.205434} {\bibfield  {journal} {\bibinfo
  {journal} {Phys. Rev. B}\ }\textbf {\bibinfo {volume} {89}},\ \bibinfo
  {pages} {205434} (\bibinfo {year} {2014})}\BibitemShut {NoStop}%
\end{thebibliography}%


\begin{thebibliography}{5}%
\makeatletter
\providecommand \@ifxundefined [1]{%
 \@ifx{#1\undefined}
}%
\providecommand \@ifnum [1]{%
 \ifnum #1\expandafter \@firstoftwo
 \else \expandafter \@secondoftwo
 \fi
}%
\providecommand \@ifx [1]{%
 \ifx #1\expandafter \@firstoftwo
 \else \expandafter \@secondoftwo
 \fi
}%
\providecommand \natexlab [1]{#1}%
\providecommand \enquote  [1]{``#1''}%
\providecommand \bibnamefont  [1]{#1}%
\providecommand \bibfnamefont [1]{#1}%
\providecommand \citenamefont [1]{#1}%
\providecommand \href@noop [0]{\@secondoftwo}%
\providecommand \href [0]{\begingroup \@sanitize@url \@href}%
\providecommand \@href[1]{\@@startlink{#1}\@@href}%
\providecommand \@@href[1]{\endgroup#1\@@endlink}%
\providecommand \@sanitize@url [0]{\catcode `\\12\catcode `\$12\catcode
  `\&12\catcode `\#12\catcode `\^12\catcode `\_12\catcode `\%12\relax}%
\providecommand \@@startlink[1]{}%
\providecommand \@@endlink[0]{}%
\providecommand \url  [0]{\begingroup\@sanitize@url \@url }%
\providecommand \@url [1]{\endgroup\@href {#1}{\urlprefix }}%
\providecommand \urlprefix  [0]{URL }%
\providecommand \Eprint [0]{\href }%
\providecommand \doibase [0]{https://doi.org/}%
\providecommand \selectlanguage [0]{\@gobble}%
\providecommand \bibinfo  [0]{\@secondoftwo}%
\providecommand \bibfield  [0]{\@secondoftwo}%
\providecommand \translation [1]{[#1]}%
\providecommand \BibitemOpen [0]{}%
\providecommand \bibitemStop [0]{}%
\providecommand \bibitemNoStop [0]{.\EOS\space}%
\providecommand \EOS [0]{\spacefactor3000\relax}%
\providecommand \BibitemShut  [1]{\csname bibitem#1\endcsname}%
\let\auto@bib@innerbib\@empty
\bibitem [{\citenamefont {Spivak}(1999)}]{SpivakII}%
  \BibitemOpen
  \bibfield  {author} {\bibinfo {author} {\bibfnamefont {M.}~\bibnamefont
  {Spivak}},\ }\href@noop {} {\emph {\bibinfo {title} {A comprehensive
  introduction to differential geometry, vol. II}}},\ \bibinfo {edition} {3rd}\
  ed.\ (\bibinfo  {publisher} {Publish or Perish, Inc.},\ \bibinfo {address}
  {Houston, Tex.},\ \bibinfo {year} {1999.})\ Chap.~\bibinfo {chapter}
  {1}\BibitemShut {NoStop}%
\bibitem [{\citenamefont {Shankar}(1994)}]{Shankar}%
  \BibitemOpen
  \bibfield  {author} {\bibinfo {author} {\bibfnamefont {R.}~\bibnamefont
  {Shankar}},\ }\href@noop {} {\emph {\bibinfo {title} {Principles of Quantum
  Mechanics}}},\ \bibinfo {edition} {2nd}\ ed.\ (\bibinfo  {publisher} {Plenum
  Press},\ \bibinfo {address} {New York},\ \bibinfo {year} {1994})\
  Chap.~\bibinfo {chapter} {5}, pp.\ \bibinfo {pages} {167--175}\BibitemShut
  {NoStop}%
\bibitem [{\citenamefont {Bardarson}(2008)}]{Bardarson}%
  \BibitemOpen
  \bibfield  {author} {\bibinfo {author} {\bibfnamefont {J.~H.}\ \bibnamefont
  {Bardarson}},\ }\href {https://doi.org/10.1088/1751-8113/41/40/405203}
  {\bibfield  {journal} {\bibinfo  {journal} {Journal of Physics A:
  Mathematical and Theoretical}\ }\textbf {\bibinfo {volume} {41}},\ \bibinfo
  {pages} {405203} (\bibinfo {year} {2008})}\BibitemShut {NoStop}%
\bibitem [{\citenamefont {Evers}\ \emph {et~al.}(2021)\citenamefont {Evers},
  \citenamefont {Aharony}, \citenamefont {Bar-Gill}, \citenamefont
  {Entin-Wohlman}, \citenamefont {Hedegård}, \citenamefont {Hod},
  \citenamefont {Jelinek}, \citenamefont {Kamieniarz}, \citenamefont
  {Lemeshko}, \citenamefont {Michaeli}, \citenamefont {Mujica}, \citenamefont
  {Naaman}, \citenamefont {Paltiel}, \citenamefont {Refaely-Abramson},
  \citenamefont {Tal}, \citenamefont {Thijssen}, \citenamefont {Thoss},
  \citenamefont {van Ruitenbeek}, \citenamefont {Venkataraman}, \citenamefont
  {Waldeck}, \citenamefont {Yan},\ and\ \citenamefont {Kronik}}]{Overview2021}%
  \BibitemOpen
  \bibfield  {author} {\bibinfo {author} {\bibfnamefont {F.}~\bibnamefont
  {Evers}}, \bibinfo {author} {\bibfnamefont {A.}~\bibnamefont {Aharony}},
  \bibinfo {author} {\bibfnamefont {N.}~\bibnamefont {Bar-Gill}}, \bibinfo
  {author} {\bibfnamefont {O.}~\bibnamefont {Entin-Wohlman}}, \bibinfo {author}
  {\bibfnamefont {P.}~\bibnamefont {Hedegård}}, \bibinfo {author}
  {\bibfnamefont {O.}~\bibnamefont {Hod}}, \bibinfo {author} {\bibfnamefont
  {P.}~\bibnamefont {Jelinek}}, \bibinfo {author} {\bibfnamefont
  {G.}~\bibnamefont {Kamieniarz}}, \bibinfo {author} {\bibfnamefont
  {M.}~\bibnamefont {Lemeshko}}, \bibinfo {author} {\bibfnamefont
  {K.}~\bibnamefont {Michaeli}}, \bibinfo {author} {\bibfnamefont
  {V.}~\bibnamefont {Mujica}}, \bibinfo {author} {\bibfnamefont
  {R.}~\bibnamefont {Naaman}}, \bibinfo {author} {\bibfnamefont
  {Y.}~\bibnamefont {Paltiel}}, \bibinfo {author} {\bibfnamefont
  {S.}~\bibnamefont {Refaely-Abramson}}, \bibinfo {author} {\bibfnamefont
  {O.}~\bibnamefont {Tal}}, \bibinfo {author} {\bibfnamefont {J.}~\bibnamefont
  {Thijssen}}, \bibinfo {author} {\bibfnamefont {M.}~\bibnamefont {Thoss}},
  \bibinfo {author} {\bibfnamefont {J.~M.}\ \bibnamefont {van Ruitenbeek}},
  \bibinfo {author} {\bibfnamefont {L.}~\bibnamefont {Venkataraman}}, \bibinfo
  {author} {\bibfnamefont {D.~H.}\ \bibnamefont {Waldeck}}, \bibinfo {author}
  {\bibfnamefont {B.}~\bibnamefont {Yan}},\ and\ \bibinfo {author}
  {\bibfnamefont {L.}~\bibnamefont {Kronik}},\ }\href@noop {} {\bibinfo {title}
  {Theory of chirality induced spin selectivity: Progress and challenges}}
  (\bibinfo {year} {2021}),\ \Eprint {https://arxiv.org/abs/2108.09998}
  {arXiv:2108.09998 [cond-mat.mtrl-sci]} \BibitemShut {NoStop}%
\bibitem [{\citenamefont {Scherer}(2010)}]{Scherer_2010}%
  \BibitemOpen
  \bibfield  {author} {\bibinfo {author} {\bibfnamefont {P.~O.~J.}\
  \bibnamefont {Scherer}},\ }\href {https://doi.org/10.1007/978-3-642-13990-1}
  {\emph {\bibinfo {title} {Computational Physics}}}\ (\bibinfo  {publisher}
  {Springer Berlin Heidelberg},\ \bibinfo {year} {2010})\ Chap.\ \bibinfo
  {chapter} {3.4}\BibitemShut {NoStop}%
\end{thebibliography}%

\end{document}


\title{Supporting Information for: Interface-Induced Conservation of Momentum Leads to Chiral-Induced Spin Selectivity}

\author{Clemens Vittmann}
\affiliation{Institut f\"ur Theoretische Physik und IQST, Albert-Einstein-Allee 11, Universit\"at Ulm, D-89081 Ulm, Germany}
\author{R. Kevin Kessing}
\affiliation{Institut f\"ur Theoretische Physik und IQST, Albert-Einstein-Allee 11, Universit\"at Ulm, D-89081 Ulm, Germany}
\author{James Lim}
\affiliation{Institut f\"ur Theoretische Physik und IQST, Albert-Einstein-Allee 11, Universit\"at Ulm, D-89081 Ulm, Germany}
\author{Susana F. Huelga}
\affiliation{Institut f\"ur Theoretische Physik und IQST, Albert-Einstein-Allee 11, Universit\"at Ulm, D-89081 Ulm, Germany}
\author{Martin B. Plenio}
\affiliation{Institut f\"ur Theoretische Physik und IQST, Albert-Einstein-Allee 11, Universit\"at Ulm, D-89081 Ulm, Germany}

\maketitle
\tableofcontents

\section{Helix model}
\subsection{Parameterization}\label{SI:sec:parametrization}
For a right-handed helix and corresponding left-handed helix, we choose parameterizations
\begin{align*}
     \mathbf{r}^\text{rh} = \begin{pmatrix} 
    R \cos\left(\frac s D \right) \\
    R \sin\left(\frac s D \right) \\
    P \frac s D
    \end{pmatrix},\qquad
    \mathbf{r}^\text{lh} = \begin{pmatrix} 
    R \cos\left(\frac s D \right) \\
    -R \sin\left(\frac s D \right) \\
    P \frac s D
    \end{pmatrix}.
\end{align*}
We will focus only on the right-handed case and drop the superscript. Curvature $\kappa$ and torsion $\tau$ of the helix are related to radius and pitch as $R =\kappa D^2$ and $P = \tau D^2$ with the length scale $D=\sqrt{R^2+P^2}$ and can be expressed via an angle $\beta = \arctan{\tau/\kappa}$ as
\begin{align*}
    \kappa = \frac{\cos\beta}{D},\quad\tau = \frac{\sin\beta}{D}.
\end{align*}
Using these definitions, the moving frame $s\mapsto \{\mathbf{e}_T(s), \mathbf{e}_N(s), \mathbf{e}_B(s)\}$ (the Frenet--Serret vectors) takes the form~\cite{SpivakII}
\begin{align}
    \mathbf{e}_T(s) = \begin{pmatrix} -\cos\beta\sin{\frac{s}{D}} \\ \cos\beta\cos{\frac{s}{D}} \\ \sin\beta \end{pmatrix},\quad 
    \mathbf{e}_N(s) = \begin{pmatrix} -\cos{\frac{s}{D}} \\ -\sin{\frac{s}{D}} \\ 0\end{pmatrix},\quad
    \mathbf{e}_B(s) = \begin{pmatrix} -\sin\beta\sin{\frac{s}{D}} \\ -\sin\beta\cos{\frac{s}{D}} \\ \cos\beta\end{pmatrix}
    \label{SI:eq:TNB}
\end{align}
and the corresponding Pauli matrices in the $\{\up,\down\}$ basis are
\begin{equation*}
    \sigma_T(s) = \begin{pmatrix}
		\sin(\beta) & -i e^{-\frac{i s}{D}} \cos(\beta) \\
		i e^{\frac{i s}{D}} \cos(\beta) & -\sin(\beta)
    \end{pmatrix}
	,\quad
	\sigma_N(s) = \begin{pmatrix}
		0 & -e^{-\frac{i s}{D}} \\
		-e^{\frac{i s}{D}} & 0
	\end{pmatrix},\quad
	\sigma_B(s) =
	\begin{pmatrix}
		\cos(\beta) & i e^{-\frac{i s}{D}} \sin(\beta) \\
		-i e^{\frac{i s}{D}} \sin(\beta) & -\cos(\beta)
	\end{pmatrix}.
\end{equation*}

\subsection{Eigenstates of the Hamiltonian}
The full form of Hamiltonian of 
\maneqref{1}
(including SOC in the bi-normal direction) is
\begin{align}
	H_h = -\frac{\hbar^2}{2m_h}{\partial_s}^2 - i\alpha_N\left(\sigma_B\partial_s -\sigma_N\frac{\tau}{2}\right) + i\alpha_B\left(\sigma_N\partial_s - \sigma_T\frac{\kappa}{2} +\sigma_B\frac{\tau}{2}\right) + U_h.
	\label{SI:eq:Hfull}
\end{align}
Defining the canonical momentum $\hat p\coloneqq m\hat v$, where the velocity operator $\hat v$ is the time derivative of the position operator $\hat s$,
\begin{align*}
    \hat p &= -\frac{i m_h}{\hbar}[\hat s, H_h] 
    = -i\hbar\partial_s + \frac{m_h}{\hbar}(\alpha_N\sigma_B-\alpha_B\sigma_N) \\
    &= -i\hbar\partial_s + \frac{m_h}{\hbar} \begin{pmatrix}
		\alpha_N \cos{\beta} & \left(\alpha_B + i \alpha_N \sin{\beta} \right) e^{-\frac{i s}{D}} \\
		\left(\alpha_B - i \alpha_N \sin{\beta} \right) e^{\frac{i s}{D}} & -\alpha_N \cos{\beta}
	\end{pmatrix} \nonumber
    ,    
\end{align*}
the Hamiltonian can be written in the manifestly self-adjoint form
\begin{align}
H_h &= \frac{\hat p^2}{2m_h} -\frac{m_h}{2\hbar^2}({\alpha_N}^2+{\alpha_B}^2) + U_h
\label{SI:eq:HamiltonianP}
\end{align}
or simply $H_h = \hat p^2/(2m_h) + U_h$ if the constant term is absorbed into $U_h$. To find the eigenvectors of $H_h$, we use a continuous basis $\ket{\xi_{k}^{\uparrow\downarrow}}$ defined as
\begin{gather}
	\ket{\xi_{k}^{\uparrow}} = \frac{1}{\sqrt{2\pi}}\ints	e^{ -\frac{i}{2}\frac{s}{D}}e^{i k s}\ket{s}\otimes \up, \nonumber\\	
	\ket{\xi_{k}^{\downarrow}}= \frac{1}{\sqrt{2\pi}}\ints	e^{\frac{i}{2}\frac{s}{D}} e^{i k s}\ket{s}\otimes \down
	\label{SI:eq:XiBasis}
\end{gather}
that satisfies $\braket{\xi_{k_1}^{s_1}}{\xi_{k_2}^{s_2}} = \delta_{s_1s_2}\delta(k_1-k_2)$. In this basis, one finds the eigenvalues
\begin{align*}
	p_{k,S} &= \hbar k + S\frac{Dm_h}{\hbar}{}\Gamma = \hbar \left(  k + \frac{S\Gamma}{2\varepsilon D}\right)
\end{align*}
of the momentum operator $\hat p$. Here, we introduced
\begin{align}
	\Gamma&\coloneqq\sqrt{\varepsilon^2 + {\varepsilon_N}^2 + {\varepsilon_B}^2 - 2 \varepsilon \varepsilon_N \cos\beta },\qquad
	\varepsilon\coloneqq \frac{\hbar^2}{2m_h D^2},\quad\varepsilon_{N,B} \coloneqq\frac{\alpha_{N,B}}{D} 
	\label{SI:eq:EpsDefs}.
\end{align}
The corresponding (unnormalized) eigenvectors are 
\begin{align*}
	\ket{k, S}	\coloneqq A_{S}\ket{\xi_{k}^{\uparrow}} + \ket{\xi_{k}^{\downarrow} }
\end{align*} with $S \in \{+,-\}$ and
\begin{align}
	A_{\pm} = \frac{ -\varepsilon+\varepsilon_N\cos\beta\pm\Gamma}{ \varepsilon_B-i\varepsilon_N\sin\beta}.
	\label{SI:eq:A}
\end{align}
Since $\hat p$ commutes with the Hamiltonian, these vectors are also eigenstates of $H_h$ and using the definitions of \eqref{SI:eq:EpsDefs}, the eigenvalues of $H_h$ are easily found as
\begin{align*}
E^{S}_{k} &= \frac{{p_{k,S}}^2}{2 m_h} - \frac{{\varepsilon_N}^2 + {\varepsilon_B}^2}{4\varepsilon} = E_{k}  + S \Delta E(k),
\end{align*}
with the energies $E_k$ and $\Delta E_k$ defined as
\begin{align*}
E_{k}&= \varepsilon\left(D^2k^2 +\frac 1 4\right)- \frac{1}{2}\varepsilon_N\cos\beta,\qquad
\Delta E_k = Dk\Gamma.
\end{align*}
Going back to the original basis, the eigenstates are
\begin{align}
	\ket{k,S} = \frac{1}{\mathcal N_{S}} \left(	A_{S} \ints e^{-\frac{i}{2}\frac{s}{D}}	e^{i k s} \ket{s}\otimes \up + 
    \ints	e^{\frac{i}{2}\frac{s}{D}}e^{i k s}\ket{s}\otimes\down\right),
    \label{SI:eq:eigenstates}
\end{align}
where we introduced a normalization constant $\mathcal{N}_{S}\coloneqq\sqrt{2\pi(1+|A_{S}|^2)}$. 

\begin{figure}
	\def\svgwidth{220pt}
	\centering
    \includegraphics[width=0.6\textwidth]{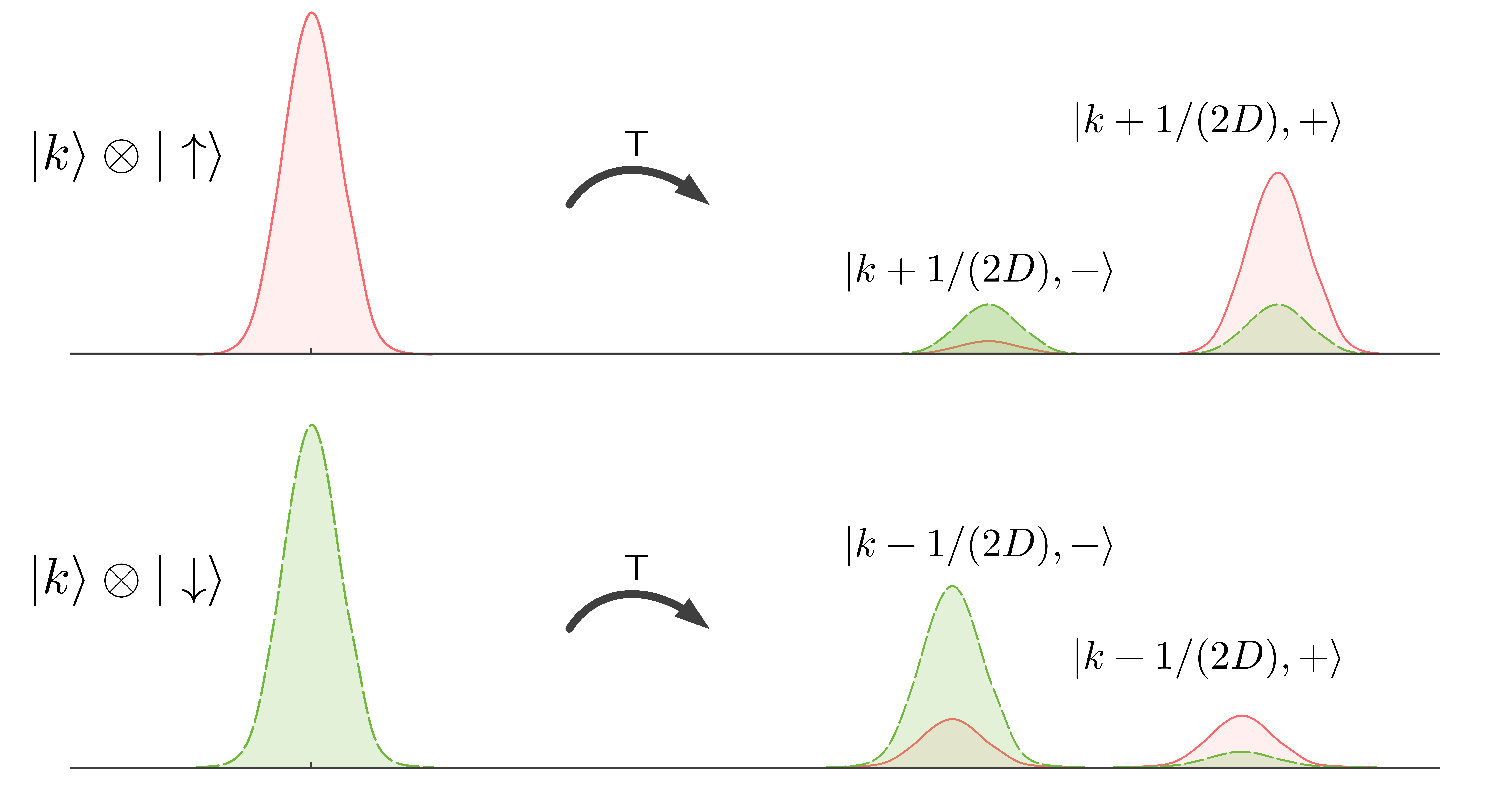}
	\caption{Wavepacket inserted into the helix with momentum $k$ and polarization $\up / \down$ splits into two oppositely polarized wavepackets moving at different velocities.}
	\label{SI:Fig:Split}
\end{figure}

\subsection{Single-channel wavepacket dynamics}
The energy gap $\Delta E(k)$ induces spatial splitting of $\up$- or $\down$-polarized wavepackets injected into the helix: The constituent $\ket{k,+}$ components of the wavepacket move faster than the $\ket{k,-}$ components, whose respective group velocities are given by $v_\pm = \hbar^{-1}\partial_kE(k)= (\hbar/m_h) k \pm D\Gamma/\hbar$, as shown in Fig.~\ref{SI:Fig:Split}. While this leads to a chirality-dependent spin current, it requires directly placing the electrons inside the helix.

Spin polarization is not obtained when placing the initial wavepacket in free space connected to a helix via a single point (e.g., in a 1D free space--helix--free space setup), with the helix Hamiltonian of \eqref{SI:eq:Hfull} and a standard free-space Hamiltonian
$H_f=-\frac{\hbar^2}{2m_e}\pdv[2]{q}$. In this setup all spin dependence can be removed from the Hamiltonian by a unitary transformation (cf.~sec.~\ref{SI:sec:gauges}).
The dynamics at each interface are then similar to the textbook problem of scattering through a rectangular potential barrier~\cite{Shankar}, where boundary conditions only require conservation of energy without any conservation of momentum. If a right-moving wavepacket is injected into the first free part, calculations of the dynamics show that no spin polarization is achieved.
This is in line with a theorem by Bardarson~\cite{Bardarson}, which can be used to prove that such a time-reversible single-channel two-terminal Hamiltonian cannot produce CISS~\cite{Overview2021}.
\subsection{Spin}
With non-zero spin-orbit coupling, the Hamiltonian eigenstates are no longer eigenstates of $\sigma_z$ but we can define a tilted spin-operator $\sigma_\text{h}$ that commutes with the Hamiltonian and has eigenvalues $S\in\{-1,1\}$. First, let
\begin{align*}
    \mathbf{e}_\text{spin} \coloneqq \frac{1}{\Gamma}\begin{pmatrix}-\varepsilon_B \\ \varepsilon_N\sin\beta \\  \varepsilon-\varepsilon_N\cos\beta\end{pmatrix},
    \qquad \sigma_h^\xi \coloneqq \bm{\sigma} \cdot \mathbf{e}_\text{spin} = \frac{1}{\Gamma} \begin{pmatrix}
    \varepsilon - \varepsilon_N \cos\beta       && -(\varepsilon_B + i\varepsilon_N\sin\beta) \\
    -(\varepsilon_B - i\varepsilon_N\sin\beta)  && -\varepsilon + \varepsilon_N\cos\beta
    \end{pmatrix}.
\end{align*}
Note that in the limit of vanishing spin-orbit coupling, $\mathbf{e}_\text{spin} \rightarrow(0,0,1)$ and $\sigma_h^\xi \rightarrow\sigma_z$. For non-zero spin-orbit coupling, the Hamiltonian eigenstates represented in basis of \eqref{SI:eq:XiBasis} are eigenvectors of $\sigma_h^\xi$ with eigenvalues $\pm 1$. Transforming to the original basis, one easily finds the explicit representation
\begin{align*}
\sigma_h = \frac{1}{\Gamma} \begin{pmatrix}
    \varepsilon - \varepsilon_N \cos\beta                       && -(\varepsilon_B + i\varepsilon_N\sin\beta)e^{-i\frac{s}{D}} \\
    -(\varepsilon_B - i\varepsilon_N\sin\beta)e^{i\frac{s}{D}}  && -\varepsilon + \varepsilon_N\cos\beta
    \end{pmatrix}
\end{align*}
which satisfies $[\sigma_h,H_h] = 0$ and $\sigma_h \psi_{k,S}(s) = -S \psi_{k,S}(s)$ for $\psi_{k,S}(s) = \braket{s}{k,S}$ and $S\in\{\pm 1\}$.
The polarization of the eigenstates of \eqref{SI:eq:eigenstates} with respect to $\sigma_z$ is
\begin{align*}
   \mathcal{P}_\pm \coloneqq \braket{k,S}{\sigma_z|k,S} = \frac{\abs{A_S}^2-1}{\abs{A_S}^2+1} = S \frac{\varepsilon_N\cos\beta - \varepsilon}{\Gamma},
\end{align*}
where we plugged in the definition from \eqref{SI:eq:A}. Writing the couplings as $\alpha_N=\cos(\gamma) \alpha$ and $\alpha_B=\sin(\gamma)\alpha$ with $\gamma\in(0,\pi/2)$, we find that in the limiting case of strong SOC:
\begin{align*}
	P_\pm^\infty\coloneqq \lim_{\alpha\rightarrow\infty} P_\pm = \pm \cos\gamma\cos\beta.
\end{align*}

\section{Gauge theory and SOC in one- and two-channel devices} \label{SI:sec:gauges}
We will now show that in the theory considered so far, all spin-dependence can be removed by a gauge transformation. First, the Legendre transform
\begin{align*}
	L = \bra{\psi}\left(i\hbar\partial_t-H_h\right)\ket{\psi}
\end{align*}
yields the Lagrangian density
\begin{align*}
	\mathcal L = i\hbar \psi^\dagger\dot\psi  + \psi^\dagger\left(\frac{\mathcal{A}\mathcal{A}}{2m_h} - U_h\right)\psi
	- \frac{1}{2m_h}\psi^\dagger\left(i\hbar \overleftarrow{\partial}_s+\mathcal{A}\right)\left(-i\hbar\overrightarrow{\partial}_s+\mathcal{A}\right)\psi
\end{align*}
which has the same equations of motion as the Hamiltonian of \eqref{SI:eq:Hfull}. Here, $\psi,\psi^\dagger$ are the spinor wavefunctions $\psi:s\mapsto\braket{s}{\psi},~\psi^\dagger: s\mapsto \braket{\psi}{s}$, $\mathcal A = \mathcal A^a T_a$ (summation over repeated indices is implicit in this section) and $\mathcal{A}^a(s)$ with $a\in\{x,y,z\}$ are defined as the coefficients in the expansion
\begin{align*}
	\mathcal{A}^aT_a = \frac{m_h}{\hbar}(\alpha_N\sigma_B-\alpha_B\sigma_N),\quad T_a = \frac{1}{2}\sigma_a.
\end{align*}
Introducing the gauge-covariant derivative $\D_s\coloneqq \partial_s+ig\mathcal{A}^aT_a$ with $g\coloneqq \hbar^{-1}$, the Lagrangian then takes the form
\begin{align}
	\mathcal{L} = i\hbar\psi^\dagger\dot\psi - \frac{\hbar^2}{2m_h}(\D_s\psi)^\dagger (\D_s\psi) - \psi^\dag U_h \psi
	+\frac{1}{2m_h}\psi^\dagger\mathcal A \mathcal A\psi.
	\label{SI:eq:LagrangianNonAbelienGaugeTheory}
\end{align}
With \eqref{SI:eq:LagrangianNonAbelienGaugeTheory}, we have re-written the Hamiltonian in a manner that resembles a non-Abelian gauge theory with gauge group $SU(2)$: Indeed, it is easy to see that, for any time-independent differentiable $\gaugetransnoarg: \mathbb{R} \to SU(2)$, a gauge transformation of the form
\begin{align*}
    \psi\mapsto \gaugetrans\psi,\quad \mathcal{A}(s)\mapsto \gaugetrans \mathcal{A}(s) \invgaugetrans + \frac{i}{g} \left(\partial_s \gaugetrans\right) \invgaugetrans,
\end{align*}
transforms $\mathcal{D}_s \psi \mapsto \gaugetrans \mathcal{D}_s \psi$ and therefore leaves first three terms in \eqref{SI:eq:LagrangianNonAbelienGaugeTheory} invariant.
Further, since the system is one-dimensional, $\mathcal A$ is pure gauge, i.e., $\mathcal A$ can be set to zero via an appropriately chosen gauge transformation.
However, the last term in \eqref{SI:eq:LagrangianNonAbelienGaugeTheory}, which is spin-independent and quadratic in the gauge field, breaks the gauge symmetry.

Nevertheless, we can define a unitary transformation 
\[U(s) = \mathcal Pe^{-\frac{i}{\hbar} \int_{s_i}^s\mathcal A(s')\dd{s'}},\]
with $s_i$ arbitrary and where $\mathcal P$ denotes path ordering with respect to $s$, such that $\partial_s U(s) = -\frac{i}{\hbar} \mathcal A(s) U(s)$. 
This transforms the Hamiltonian $H_h$ of \eqref{SI:eq:HamiltonianP} into a spin-independent form:
\begin{align*}
U H U^{-1} = -\frac{\hbar^2}{2m_h} {\partial_s}^2 - \frac{m_h}{2\hbar^2}({\alpha_N}^2+{\alpha_B}^2) + U_h,
\end{align*}
where the first term is the standard kinetic energy of a free particle and the remaining terms are constant and therefore physically irrelevant energy shifts.
This demonstrates that the Hamiltonian of the helical path with SOC is unitarily equivalent to a quasi-free particle, and therefore
SOC alone can not lead to spin polarization effects. 
However, this changes in a coupled-waveguide model as considered in the main text: The states then 
take the form $\psi(s,q) = \psi_h(s) + \psi_f(q)$ with $\psi_h\perp\psi_f$ and the Hamiltonian is $H = H_h + H_f + V$ with $V$ the coupling between the two channels. Then, the unitary transformation that removes the spin-orbit coupling from the helix Hamiltonian is
\begin{align*}
    \mathcal U(s,q) = U(s)P_s + P_q,
\end{align*}
where $P_s, P_q$ are the projectors $P_s\psi(s,q) = \psi_h(s)$ and $P_q\psi(s,q) = \psi_f(q)$. While $\mathcal U$ indeed renders the helix Hamiltonian spin-independent while leaving the free-channel part unchanged, the coupling $V(s,q) = f(s,q)(\dyad{s}{q} + \dyad{q}{s})$ is transformed as
\begin{align*}
    V'(s,q) &= (U(s)P_s+P_q)V(s,q)(P_sU^{-1}(s)+P_q)\\
    &= f(s,q) \left[ U(s)\dyad{s}{q} + \dyad{q}{s}U^{-1}(s) \right] 
\end{align*}
which is spin-dependent as long as the coupling is extended over a region. Hence, the spin-orbit coupling cannot, in general, be removed by a unitary transformation. However, for a point-like coupling at position $(s_0,x_0)$, one has $f(s,q)= f_0\delta(s-s_0)\delta(q-q_0)$ and
\begin{align*}
    V'=\int_{\mathbb{R}^2}\dd{s}\dd{q}V'(s,q) = f_0 \left[ U(s_0)\dyad{s_0}{q_0} + \dyad{q_0}{s_0}U^{-1}(s_0) \right].
\end{align*}
Since we are free to choose $U(s)=\mathcal P e^{-i g\int_{s_i}^s\mathcal A(s')\dd{s'} + c}$ with $c\in \mathfrak{su}(2)$, one can take $U(s_0) = \mathbbm{1}$ and hence,
$V' = f_0(\dyad{s_0}{q_0}+\dyad{q_0}{s_0})$. Thus, we have shown that in the limit of a point-like coupling, the Hamiltonian of the coupled two-channel system is unitarily equivalent to one that is spin-independent and hence, cannot produce spin polarization.

\section{Numerics} \label{SI:sec:numerics}
To compute the time evolution numerically, we employ the fourth-order Runge-Kutta method. Given a spinor wavefunction $[a,b]\ni s\mapsto (\psi_\uparrow(s),\psi_\downarrow(s))$, the interval $[a,b]$ is divided into $N$ discrete points $(s_1,...,s_N)$ with distance $\Delta s\coloneqq s_{j+1}-s_j$. Accordingly, a state $\psi$ is represented by $2N$ complex numbers 
and the Hamiltonian is a $(2N\times2N)$-matrix. In our simulations, we choose $N$ such that $\Delta s$ is on the order of $\SI{0.01}{\nano\meter}$. 
To obtain the 
state $\psi^t$ at some later time $t$, we divide the time interval $t$ into $M$ small time segments $\Delta t$ (on the order of $\SI{0.01}{\femto\second}$ in our simulations). Then, four $2N$-component vectors $k_1,...k_4$ are defined as\footnote{Note that the $k_1$ through $k_4$ defined here are unrelated to the wavenumbers and crossing points mentioned in the main text.}
\begin{equation*}
    k_1 = (H \cdot\psi)\Delta t,\qquad
    k_2 = (H \cdot k_1)\frac{\Delta t}{2},\qquad
    k_3 = (H \cdot k_2)\frac{\Delta t}{3},\qquad
    k_4 = (H \cdot k_3)\frac{\Delta t}{4}
\end{equation*}
and $\psi^{\Delta t} = \psi + k_1 + k_2 + k_3 + \mathcal{O}\fleft((\Delta t)^5\fright)$. Applying this scheme $M$ times then results in the state $\psi^t$ at time $t=M\Delta t$.

Since the Hamiltonian contains derivatives $\partial_s$ and ${\partial_s}^2$, we also need to discretize the first and second derivative operators. This is done using fourth-order finite difference schemes as \cite{Scherer_2010}
\begin{align*}
    \eval{\pdv{\psi}{s}}_{s_j} &\approx \frac{1}{12}\psi_{j-2} -\frac{2}{3}\psi_{j-1} + \frac{2}{3}\psi_{j+1}-\frac{1}{12}\psi_{j+2},\\
    \eval{\pdv[2]{\psi}{s}}_{s_j} &\approx -\frac{1}{12}\psi_{j-2} +\frac{4}{3}\psi_{j-1} - \frac{5}{2}\psi_j + \frac{4}{3}\psi_{j+1}-\frac{1}{12}\psi_{j+2}.
\end{align*}

\section{Extrapolation to and saturation behavior of multiple scatterings using transfer matrices}
In Fig.~2(d) of the main text, we showed that for multiple concatenated scattering events, the spin polarization of the transmitted free wavepacket is amplified. Here, describe how we estimate the behavior of these multiple-scattering events from that of a single scattering, and we explaing why the numerical simulations indicate that the spin polarization after $n$ events, $P_n$, generally levels off at values less than \SI{100}{\percent}.

We illustrate how this behavior arises using a transfer-matrix-like approach that is valid in the narrow-bandwith (quasi-monochromatic) limit:
Due to conservation of energy, an incoming free-waveguide state of some definite $k_0 > 0$ will scatter to free-waveguide states of the same $\abs{k} = k_0$, as well as to helical-waveguide states of a different but nevertheless definite $k_\pm$ satisfying $E\fleft(k_\pm, \pm\fright) = \hbar^2 {k_0}^2 / (2m_e)$. Therefore, assuming negligible reflection to negative-$k$ components, we can represent the combined free- and helical-waveguide wavefunction as a simple four-tuple:
\[\ket{\phi} = f_\uparrow\ket{k_0}\ket{\uparrow} + f_\downarrow\ket{k_0}\ket{\downarrow} + h_+\ket{k_+, +} + h_-\ket{k_-, -} 
\iff
\phi = \begin{pmatrix} f_\uparrow \\ f_\downarrow \\ h_+ \\ h_- \end{pmatrix},\]
where the first two components, $f_\uparrow$ and $f_\downarrow$, form the free-waveguide spinor and the latter two, $h_+$ and $h_-$, the helical-waveguide spinor.
In this monochromatic limit, a state that was initially prepared in some ${\phi}_0$ and then undergoes scattering in an interaction region is given exactly by $U{\phi}_0$, where $U$ is a unitary $4\times 4$ matrix which is theoretically computable from the propagator $U = e^{-iHt/\hbar}$ of the system. 
This is essentially equivalent to a beamsplitter description of the system in which the spin states of the electron are analogous to the polarization direction of light.
Note that the matrix $U$ can be thought of as consisting of four $2\times 2$ matrices,
\[U = \begin{pmatrix} U_{ff} & U_{fh}\\ U_{hf} & U_{hh}\end{pmatrix},\]
where the block $U_{ff}$ ($U_{hh}$) describes the relationship between the incoming and outgoing free-waveguide (helical-waveguide) components, whereas the off-diagonal blocks quantify the dynamics of transfer from helical to free waveguide and vice versa.

We can now further simplify this approach by realizing that for the purposes of calculating the free-waveguide spin polarization, we are only interested in the free-waveguide components $f_\uparrow$ and $f_\downarrow$. 
In addition, the incoming state to each scattering event also consists, by assumption, only of such free-waveguide components. 
Therefore, we can project onto the free waveguide before and acting with $U$, such that in effect only the upper-left $2\times 2$ block $U_{hh}$ is of interest to our analysis and we can reduce our system from four to two dimensions. 
However, note that this matrix, which we call $U_{hh} = M$ from now on, is no longer unitary, which is to be expected as the scattering events are not norm-conserving from the perspective of the free-waveguide state.

In this formalism, $n$ repeated scatterings of a pure state ${\phi}$ are easily described as $M^n {\phi}$, and thus we obtain the average polarization of an unpolarized stochastic mixture of states, $\rho_0 = \frac{1}{2} \mathbbm{1}_{2\times 2}$, after $n$ scatterings as
\begin{equation}
P_n = \frac{\Tr{\sigma_z \rho_n}}{\Tr{\rho_n}} 
= \frac{\Tr{\sigma_z M^n \rho_0 \left(M^\dag\right)^n}}{\Tr{M^n \rho_0 \left(M^\dag\right)^n}} 
= \frac{\Tr{\left(M^\dag\right)^n \sigma_z M^n}}{\Tr{\left(M^\dag\right)^n M^n}} 
. \label{SI:eq:Pn_trace}
\end{equation}
In Fig.~2(d) of the main text, we show the predictions that are obtained from \eqref{SI:eq:Pn_trace} by approximating the matrix elements of $M$ from the populations after a single scattering as:
\begin{equation}
M \approx \begin{pmatrix}\sqrt{p_{\uparrow\uparrow}} & \sqrt{p_{\uparrow\downarrow}} \\ \sqrt{p_{\downarrow\uparrow}} & \sqrt{p_{\downarrow\downarrow}} \end{pmatrix} \qq{where} p_{\sigma_j \sigma_i} = \int_\text{free} \abs{\Big[\bra{\sigma_j} \otimes \bra{q}\Big] \ket{\psi_i^{\sigma_i}(t)}}^2 \dd{q},\label{SI:eq:M}
\end{equation}
where $\ket{\psi_i^{\sigma_i}(t)}$ is the state 
(\maneqref{5}) 
with polarization $\sigma_i$ after passing through the interaction region ($t\to\infty$), and where the integration is only over the free-waveguide part of the system. The quasi-monochromatic limit is then applicable in the limit of large $\zeta$.

We can also show analytically that the many-scattering limit polarization is, in general, less than \SI{100}{\percent}, i.e., $\abs{\lim_{n\to\infty} P_n} < 1$. 
For this analysis, we assume $M$ to be real and symmetric, as is the case in \eqref{SI:eq:M}.\footnote{It is evident from the second-order Dyson series term that the two probabilities $p_{\uparrow\downarrow}$ and $p_{\downarrow\uparrow}$ are equal, as the two Feynman diagrams are composed of identical vertices in opposite order, which ensures that our approximate representation of $M$ is symmetric.} 
Then we can diagonalize $M$ as 
\begin{equation}
    M = {U_\theta}^{-1} \Lambda U_\theta \qq{with} U_\theta = \begin{pmatrix} \cos{\theta} & \sin{\theta}\\ -\sin{\theta} & \cos{\theta} \end{pmatrix} \qand \Lambda = \begin{pmatrix} \lambda_1 & 0 \\ 0 & \lambda_2 \end{pmatrix}, \qq{where} \tan(2\theta) = \frac{2 \sqrt{p_{\downarrow\uparrow}}}{\sqrt{p_{\uparrow\uparrow}} - \sqrt{p_{\downarrow\downarrow}}},
    \label{SI:eq:2x2diagonal}
\end{equation}
which allows us to easily compute $M^n = {U_\theta}^{-1} \Lambda^n U_\theta$. Inserting this into \eqref{SI:eq:Pn_trace}, one finds 
after a straightforward calculation 
that
\[P_n = \frac{\Tr{M^{2n} \sigma_z}}{\Tr{M^{2n}}} 
= \frac{\eta^n - 1}{\eta^n + 1} \cos(2\theta) \qq{where} \eta \coloneqq \left(\frac{\lambda_1}{\lambda_2}\right)^2 \geq 0.\]
It is then easy to see that the many-scattering limit is
\[\lim_{n\to\infty} P_n = P_\infty = \begin{cases} \cos(2\theta) &\qif* \eta > 1,\\ -\cos(2\theta) &\qif* \eta < 1,\\ 0 &\qif* \eta = 1. \end{cases}\]
We can interpret this result by first considering the case in which spin-flip processes occur with very low probabilities $p_{\downarrow\uparrow}$, such that $\theta \approx 0$ according to \eqref{SI:eq:2x2diagonal}: Then $\eta \approx p_{\uparrow\uparrow}/p_{\downarrow\downarrow}$ and the limiting polarization will tend toward $+1$ if $\up$ states are more likely to be transmitted unabsorbed and $-1$ if $\down$ states are more readily transmitted, or $0$ if both are transmitted with equal probabilities, as one might naively expect from such a saturation. However, once spin-flip processes are introduced, $\theta \neq 0$ and the limiting polarization $\abs{P_\infty} < 1$, which immediately demonstrates how second-order processes (in the coupling strength $V_0$) can cause the polarization to saturate to values of less than \SI{100}{\percent}.

\section{Wavefunction overlap model}
\subsection{Overlap between Gaussian functions in a wedge geometry}
\subsubsection{Definitions and geometry}
Let the local coordinate system of the helical waveguide be $(\hat{n}_1, \hat{n}_2, \hat{s})$ and that of the straight waveguide $(\hat{x}_1, \hat{x}_2, \hat{z})$. The axes $(\hat{n}_1, \hat{n}_2, \hat{s})$ correspond to $(\hat{B}, -\hat{N}, \hat{T})$ in the Frenet--Serret frame.
The helical waveguide shall be discretized into wedge-like sections that are bounded by two non-parallel planes. The straight waveguide shall be discretized into simple discs bounded by parallel planes.

We begin by calculating the interaction in the case of zero thickness, i.e., the case of two-dimensional planes. Then the interaction between the two segments on the individual waveguides is described by an integral on the intersection of these two planes, which is a line integral (or a 2D integral in a degenerate geometry). This line integral follows the line of nodes given by the intersection between the $x_1$-$x_2$ and $n_1$-$n_2$ planes.

We begin with the situation in which the two coordinate systems are coincident, 
($\hat{n}_1 \parallel \hat{x}_1$ and $\hat{n}_2 \parallel \hat{x}_2$ and $\hat{s} \parallel \hat{z}$),
where the $z$ system shall remain fixed. We then consider three elemental rotations about the $s$ system, first about the $\hat{z}$ (or $\hat{s}$) axis, then about the $\hat{n}_1$ axis and finally about the $\hat{s}$ axis again -- these rotations correspond to the common $z$-$x'$-$z''$ Euler angle convention. We follow the usual Euler angle convention of calling the angles of these three rotations $\alpha$, $\beta$, $\gamma$:\footnote{Note that the quantities $\alpha$, $\beta$, $\gamma$ of sec.~\ref{SI:sec:parametrization} are unrelated to these Euler angles.} $\alpha$ is then the angle between the $\hat{x}_1$ axis and the line of nodes, $\beta$ is the angle between the $\hat{z}$ axis and the $\hat{s}$ axis and $\gamma$ is the angle between the line of nodes and the $\hat{n}_1$ axis.

First, we immediately notice that if $\beta = 0$, then the line of nodes is degenerate and becomes a plane instead of a line. Therefore, we consider these two cases separately, beginning with the more general non-degenerate case, $\beta \neq 0$. In this case, the overlap between two slices of Gaussians of infinitesimal thickness is given by the line of nodes between the two coordinate systems.
Furthermore, let the position of the helical segment's origin relative to the straight origin be given in cylindrical coordinates, by a radial distance $d_r$, a height distance $d_z$ and an angle $\varphi$.
\begin{figure}
\centering
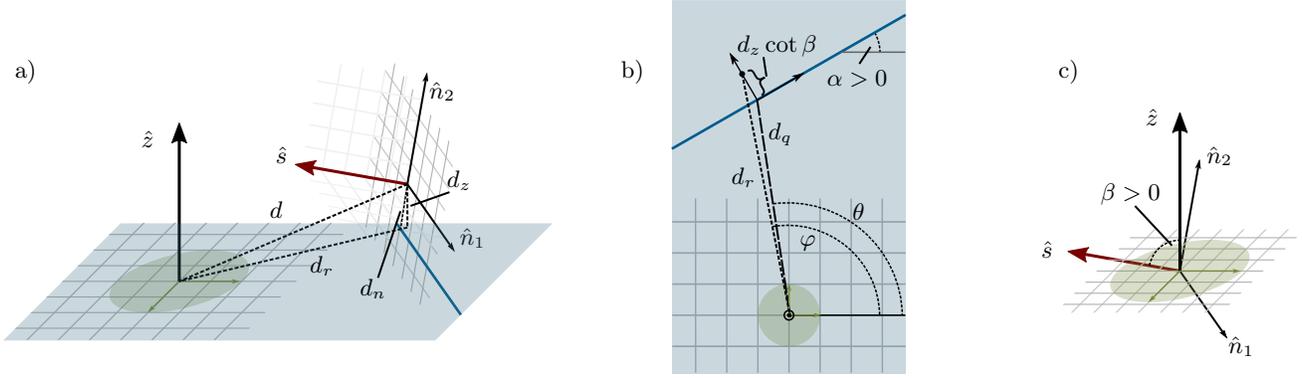
\caption{The geometry considered in the overlap-based coupling matrix calculations, showing the arrangement of the various angles involved: The two \enquote{principal} axes are $\hat{s}$ and $\hat{z}$, the tangent axis (and thus local direction of electron propagation) of the helical and straight waveguide, respectively. a) and b) show a side view and top-down view of the same configuration, respectively, whereas c) shows the two coordinate systems superimposed.
}
\label{SI:fig:geometry}
\end{figure}

In the non-degenerate case (Fig.~\ref{SI:fig:geometry}), let the line of nodes be parametrized as a function of a variable $t$. Let the point $t=0$ be the point on the line of nodes that is closest to the $s$ origin, and call this distance $d_n \coloneqq d_z / \sin{\beta}$. 
Further, define
\begin{align*}
{d_q}^2 &= {d_r}^2 + \left(d_z \cot{\beta}\right)^2 - 2d_r d_z \cot{\beta} \sin(\varphi - \alpha)
\end{align*}
and
\[\theta \coloneqq \atantwo\fleft(d_r \sin{\varphi} - d_z \cot{\beta} \cos{\alpha}, d_r \cos{\varphi} + d_z \cot{\beta} \sin{\alpha}\fright).\]
Then the line of nodes in the helical system as a function of $t$ is
\[\begin{pmatrix}n_1\\n_2\end{pmatrix} = \ell_n(t) \coloneqq \begin{pmatrix} -d_n\sin{\gamma}\\ -d_n \cos{\gamma} \end{pmatrix} + t\begin{pmatrix} \cos{\gamma}\\ -\sin{\gamma} \end{pmatrix}\]
and in the Cartesian system:
\[
\begin{pmatrix}x_1\\x_2\end{pmatrix} = \ell_x  (t) \coloneqq \underbrace{d_r \begin{pmatrix}\cos{\varphi}\\ \sin{\varphi} \end{pmatrix} + d_z \cot{\beta} \begin{pmatrix}\sin{\alpha}\\ -\cos{\alpha} \end{pmatrix}}_{\eqqcolon d_q \begin{pmatrix}\cos{\theta}\\ \sin{\theta} \end{pmatrix}} + t\begin{pmatrix} \cos{\alpha}\\ \sin{\alpha} \end{pmatrix}.
\]
Thus, we know that the overlap integral in question will depend on the integral of the two wavefunctions with the $x$ and $n$ components given by the parametrization of the line of nodes:
\[
\int_\mathbb{R} \phi^\ast\fleft(\ell_n(t),\, s\fright)
     \psi\fleft(\ell_x(t),\, z\fright) \dd{t}.
\]

Furthermore, we need to consider the fact that the standard Lebesgue integration measure weights every point equally, which is not the behavior we want here. Instead, in order to capture the wedge structure of the helical segments, we need to linearly increase the weight of the integration measure in one direction and let it decrease to zero in the opposite direction. Let $\hat{n}_2$ be the direction impacted by this scaling (and therefore $\hat{n}_1$ must always be perpendicular to the wedge direction and extend over all of $\mathbb{R}$):
\begin{equation*}
\dd{n_2} \longmapsto  f(n_2) \dd{n_2} \qq{with} f(n_2) = \begin{cases} 1 + \frac{n_2}{R} & \text{if } n_2 > -R, \\ 0 & \text{else}. \end{cases}
\end{equation*}
Thus, we obtain
\[
I(s, z) = \int_\mathbb{R} f\fleft(-d_n \cos{\gamma} - t\sin{\gamma}\fright)
\phi^\ast\fleft(\ell_n(t),\, s\fright)
     \psi\fleft(\ell_x(t),\, z\fright) \dd{t},
\]
where the argument of $f$ is the $n_2$ component of $\ell_n(t)$.

Finally, we need to integrate this line integral over an area, since we want each element of our discretization of the waveguides to cover a finite volume:
\[I(S_i, Z_j) = \int_{s \in S_i} \int_{s \in Z_j} I(s, z) \dd{s} \dd{z}.\]
As an approximation for sufficiently small discretization elements, the geometry can be assumed to be constant over the discretization element, allowing us to approximate
\[I(S_i, Z_j) \approx \abs{S_i} \abs{Z_j} I(s_i, z_j),\]
where $s_i$ and $z_j$ are some representative values of $S_i$ and $Z_j$, respectively. We choose the value at the center of $S_i$ as $s_i$ (and similarly for $Z_j$ and $z_j$).

\subsubsection{Inserting Gaussians and evaluating the integral for \texorpdfstring{$\beta \neq 0$}{beta non-zero}}
We can now insert the Gaussian wavefunctions,
\[\phi\fleft(n_1, n_2, s \in S_i\fright) = \frac{1}{\sqrt{\abs{S_i} \pi {\sigma_n}^2}} e^{-\frac{{n_1}^2 + {n_2}^2}{2{\sigma_n}^2}}\]
(which have the same functional form for $\psi(x_1, x_2, z)$), along with the line-of-nodes parametrization to obtain:
\[
I(S_i, Z_j) \approx  K \int_\mathbb{R} 
f\fleft(-d_n \cos{\gamma} - t\sin{\gamma}\fright)  \exp\fleft\{-\frac{1}{2\sigma^2} \left(t + \frac{d_q \sigma^2 \cos(\theta -\alpha)}{{\sigma_x}^2}\right)^2 \fright\}
\dd{t}
 \]
with
\begin{equation}
K \coloneqq \frac{\sqrt{\abs{S_i} \abs{Z_j}}}{\pi \sigma_s \sigma_x} \exp\fleft\{-\frac{{d_n}^2}{2{\sigma_n}^2} - \frac{{d_q}^2}{2{\sigma_x}^2} \left(1 - \frac{\sigma^2 \cos^2(\theta -\alpha)}{{\sigma_x}^2}\right)\fright\}
\label{SI:eq:K}
\end{equation}
and
\begin{equation}
\frac{1}{\sigma^2} \coloneqq  \frac{1}{{\sigma_x}^2} + \frac{1}{{\sigma_n}^2} \quad \iff \quad \sigma^2 = \frac{{\sigma_x}^2 {\sigma_n}^2}{{\sigma_x}^2 + {\sigma_n}^2}.
\label{SI:eq:sigma}
\end{equation}
Moreover, the scaling function $f$ is found to have the following form here:
\begin{equation}
f(n_2) = \begin{cases} 
    1 - \frac{d_n}{R} \cos{\gamma}                  & \qif* \gamma\bmod \pi = 0    \qand d_n \cos{\gamma} < R, \\
    1 - \frac{t\sin{\gamma} + d_n \cos{\gamma}}{R}   & \qif* \gamma\bmod 2\pi < \pi \qand t < \frac{R}{\sin{\gamma}} - d_n \cot{\gamma},\\
    1 - \frac{t\sin{\gamma} + d_n \cos{\gamma}}{R}   & \qif* \gamma\bmod 2\pi > \pi \qand t > \frac{R}{\sin{\gamma}} - d_n \cot{\gamma},\\
    0                                               & \text{else}.
\end{cases}
\label{SI:eq:cases}
\end{equation}
Therefore, we separate the evaluation of the integral into three cases.

\paragraph{\texorpdfstring{$\gamma \bmod 2\pi < \pi$}{gamma mod (2pi) < pi}}
In the case in which the line of nodes is not parallel to the $\hat{n}_1$ axis \emph{and} $\sin{\gamma} > 0$, there is a variable scaling component given by $f$, and the integral has an upper bound that is determined by $f$:
\[
I(S_i, Z_j) \approx \frac{K}{R} \int_{-\infty}^{\frac{R}{\sin{\gamma}} - d_n \cot{\gamma}}
\left(R -d_n \cos{\gamma} - t\sin{\gamma}\right)  \\
 \exp\fleft\{-\frac{1}{2\sigma^2} \left(t + \frac{d_q \sigma^2 \cos(\theta -\alpha)}{{\sigma_x}^2}\right)^2 \fright\}
\dd{t}
\]
with $K$ as given in \eqref{SI:eq:K}.
Evaluating the first two terms is straightforward by using the definition of the error function $\erf$.
Using partial integration and defining $\lambda \coloneqq \frac{d_q \sigma^2 \cos(\theta - \alpha)}{{\sigma_x}^2}$, the third term is equal to
\begin{multline*}
-\frac{K\sin{\gamma}}{R} \int_{-\infty}^{\frac{R}{\sin{\gamma}} - d_n \cot{\gamma}}
t \exp\fleft\{-\frac{1}{2\sigma^2} \left(t + \lambda \right)^2\fright\}
\dd{t}\\
= -\frac{K\sin{\gamma}}{R} \Bigg[-\sigma^2 \exp\fleft\{-\frac{1}{2\sigma^2} \left(\frac{R}{\sin{\gamma}} - d_n \cot{\gamma} + \lambda \right)^2\fright\}
    - \lambda \int_{-\infty}^{\frac{R}{\sin{\gamma}} - d_n \cot{\gamma}} e^{-\frac{1}{2\sigma^2} \left(t + \lambda\right)^2}
\dd{t}\Bigg].
\end{multline*}
Thus we obtain
\begin{equation}
I(S_i, Z_j) \approx \frac{\sigma^2 K \sin{\gamma}}{R} \left(\mathcal{U} \sqrt{\pi} 
\left[\erf\fleft(\mathcal{U}\fright) + 1\right]
+ e^{-\mathcal{U}^2} \right), 
\label{SI:eq:I_sinpositive}
\end{equation}
where another shorthand was defined:
\[\mathcal{U} \coloneqq \frac{\frac{R}{\sin{\gamma}} - d_n \cot{\gamma} + \lambda }{\sqrt{2\sigma^2}} = 
\frac{\frac{R}{\sin{\gamma}} - d_z \frac{\cot{\gamma}}{\sin{\beta}} + \frac{\sigma^2 d_q \cos(\theta - \alpha)}{{\sigma_x}^2} }{\sqrt{2\sigma^2}}.\]

\paragraph{\texorpdfstring{$\gamma \bmod 2\pi > \pi$}{gamma mod (2pi) > pi}}
The second case is almost identical to the first, but applies when $\sin{\gamma} < 0$. Then the sign changes and the integral's upper bound becomes the lower bound instead:
\[
I(S_i, Z_j) \approx \frac{K}{R} \int_{\frac{R}{\sin{\gamma}} - d_n \cot{\gamma}}^{\infty}
\left(R -d_n \cos{\gamma} - t\sin{\gamma}\right)
 \exp\fleft\{-\frac{1}{2\sigma^2} \left(t + \frac{d_q \sigma^2 \cos(\theta -\alpha)}{{\sigma_x}^2}\right)^2 \fright\}
\dd{t}.
\]
Using the same procedure as above:
\begin{multline*}
-\frac{K\sin{\gamma}}{R} \int_{\frac{R}{\sin{\gamma}} - d_n \cot{\gamma}}^{\infty}
t \exp\fleft\{-\frac{1}{2\sigma^2} \left(t + \lambda \right)^2\fright\}
\dd{t}\\
= -\frac{K\sin{\gamma}}{R} \Bigg[\sigma^2 \exp\fleft\{-\frac{1}{2\sigma^2} \left(\frac{R}{\sin{\gamma}} - d_n \cot{\gamma} + \lambda \right)^2\fright\}
    - \lambda \int_{\frac{R}{\sin{\gamma}} - d_n \cot{\gamma}}^{\infty} e^{-\frac{1}{2\sigma^2} \left(t + \lambda\right)^2}
\dd{t}\Bigg].
\end{multline*}
By modifying the shorthand $\mathcal{U}$ to be defined in the following, more general way,
\[\mathcal{U} \coloneqq \frac{\frac{R}{\sin{\gamma}} - d_n \cot{\gamma} + \lambda }{\sgn\fleft(\sin{\gamma}\fright)\sqrt{2\sigma^2}} = 
\frac{\frac{R}{\sin{\gamma}} - d_z \frac{\cot{\gamma}}{\sin{\beta}} + \frac{\sigma^2 d_q \cos(\theta - \alpha)}{{\sigma_x}^2} }{\sgn\fleft(\sin{\gamma}\fright) \sqrt{2\sigma^2}},\]
we can define the general equation for both $\sin{\gamma} < 0$ and $\sin{\gamma} > 0$:
\begin{equation}
I(S_i, Z_j) \approx \frac{\sigma^2 K \abs{\sin{\gamma}}}{R} \left(\mathcal{U} \sqrt{\pi} 
\left[\erf\fleft(\mathcal{U}\fright) + 1\right]
+ e^{-\mathcal{U}^2} \right) 
\label{SI:eq:I_general}
\end{equation} 
with $\sigma$ as defined in \eqref{SI:eq:sigma} and $K$ as defined in \eqref{SI:eq:K}. This contains \eqref{SI:eq:I_sinpositive} as a special case.

\paragraph{\texorpdfstring{$\gamma\bmod\pi = 0$}{gamma = 0}}
In the case in which the line of nodes \emph{is} parallel to the $\hat{n}_1$ axis (i.e., $\sin{\gamma} = 0$), the line of nodes is infinitely long and the scaling is constant, as given by \eqref{SI:eq:cases}:
\begin{equation}
I(S_i, Z_j) \approx 
    \begin{cases}
    K \left(1 - \frac{d_n}{R} \cos{\gamma}\right) \sqrt{2\pi\sigma^2} &\qif* d_n \cos{\gamma} < R,\\
    0    &\text{else}.
    \end{cases}
    \label{SI:eq:Igamma0}
\end{equation}

\subsubsection{Evaluating the integral for \texorpdfstring{$\beta = 0$}{beta = 0}}
For the parallel (degenerate) case, $\beta = 0$, the situation is slightly different, since the intersection of the integration sets is then a plane instead of a line. The case $\beta = 0$, intuitively, means that the helical segment is pointing in the same direction as the straight waveguide.  

Due to the relatively simple geometry in this case, it is easier to compute the full 3D integral here. 
Since $\beta = 0$ is a fringe case that rarely occurs, we do not cover its (rather straightforward) derivation in detail. The final result in this case is
\begin{align*}
I_0(S_i, Z_j)   
    &= \frac{\abs{S_i \cap Z_j}}{\sqrt{\abs{S_i}\abs{Z_j}}} \sqrt{\frac{2\sigma^6}{\pi R^2 {\sigma_n}^2 {\sigma_x}^2}} e^{-\frac{{d_r}^2}{2\left({\sigma_n}^2 + {\sigma_x}^2\right)} } \left( \mathcal{U}' \sqrt{\pi} \left[\erf\fleft(\mathcal{U}'\fright) + 1\right] + e^{-\left(\mathcal{U}'\right)^2}\right),
\end{align*}
where
\[\mathcal{U}' \coloneqq \frac{1}{\sqrt{2\sigma^2}} \left(R - \frac{d_r \sigma^2 \sin(\varphi -\alpha)}{{\sigma_x}^2}\right),\]
which takes on an analogous role to $\mathcal{U}$ defined for $\beta \neq 0$.

\subsection{Including helix parametrizations}
As stated above, we use Euler angles corresponding to a $z$-$x'$-$z''$ sequence of intrinsic active rotations. That is, the rotation matrix corresponding to our Euler angles $\alpha$, $\beta$, $\gamma$ is given by
\begin{multline}
\mathfrak{R} = 
\begin{pmatrix} \cos{\alpha}    & -\sin{\alpha} & 0\\ 
                \sin{\alpha}    & \cos{\alpha} & 0\\ 
                0               & 0             & 1 \end{pmatrix}
\begin{pmatrix} 1               & 0             & 0\\ 
                0               & \cos{\beta}   & -\sin{\beta}\\ 
                0               & \sin{\beta}   & \cos{\beta} \end{pmatrix}
\begin{pmatrix} \cos{\gamma}    & -\sin{\gamma} & 0\\ 
                \sin{\gamma}    & \cos{\gamma} & 0\\ 
                0               & 0             & 1 \end{pmatrix} 
\\=
\begin{pmatrix}   \cos{\alpha} \cos{\gamma} - \cos{\beta}\sin{\alpha} \sin{\gamma}    & -\cos{\alpha} \sin{\gamma} - \cos{\beta}\cos{\gamma}\sin{\alpha}  & \sin{\alpha} \sin{\beta}\\ 
                    \cos{\gamma} \sin{\alpha} + \cos{\alpha} \cos{\beta} \sin{\gamma}   & \cos{\alpha}\cos{\beta}\cos{\gamma} - \sin{\alpha}\sin{\gamma}    & -\cos{\alpha}\sin{\beta}\\
                    \sin{\beta} \sin{\gamma}                             & \cos{\gamma}\sin{\beta}                            & \cos{\beta}
\end{pmatrix}. \label{SI:eq:rotmatrix}
\end{multline}
We explicitly use this rotation matrix $\mathfrak{R}$ in the following.
\subsubsection{Geometrical parameters for the base case: Helix axis parallel to straight waveguide}
From \eqref{SI:eq:rotmatrix}, we can immediately derive
\begin{equation*}
\alpha  = \atantwo\fleft(\mathfrak{R}_{13}, -\mathfrak{R}_{23}\fright), \qquad
\beta   = \arccos(\mathfrak{R}_{33}), \qquad
\gamma  = \atantwo\fleft(\mathfrak{R}_{31}, \mathfrak{R}_{32}\fright).
\end{equation*}
Now, recall that each matrix element $\mathfrak{R}_{ij}$ of $\mathfrak{R}$ (a change-of-basis matrix) can be understood as the $i$-th component of the $j$-th transformed basis vector, such that
\begin{equation}
\alpha  = \atantwo\fleft(\hat{x} \cdot \hat{s}, -\hat{y} \cdot \hat{s}\fright), \qquad
\beta   = \arccos(\hat{z} \cdot \hat{s}), \qquad
\gamma  = \atantwo\fleft(\hat{z} \cdot \hat{n}_1, \hat{z} \cdot \hat{n}_2\fright).
\label{SI:eq:Eulerangles}
\end{equation}
Therefore, using the Frenet--Serret frame of \eqref{SI:eq:TNB} and the identification $(\hat{n}_1, \hat{n}_2, \hat{s}) = (\hat{B}, -\hat{N}, \hat{T})$, we find the Euler angles to be
\begin{equation*}
\alpha  = \atantwo\fleft(-R\sin(\frac{s}{D}), -R\cos(\frac{s}{D})\fright), \qquad
\beta   = \arccos(\frac{P}{D}), \qquad
\gamma  =  \frac{\pi}{2} \sgn\{R\}.
\end{equation*}
Having established the angular geometry, we proceed to the absolute configuration. Let the helix be shifted in the $x$-$y$ plane by $(x_0, y_0, 0)$. 
We also need the radial distance $d_r$ as a function of $s = zD/P$,
\begin{align*}
d_r & 
    = \sqrt{{x_0}^2 + {y_0}^2 + R^2 + 2R\left(x_0 \cos(\frac{s}{D}) + y_0 \sin(\frac{s}{D})\right)},
\end{align*}
and the angle $\varphi$,
\begin{align*}
\varphi &= \atantwo\fleft(y_0 + R\sin(\frac{s}{D}), x_0 + R\cos(\frac{s}{D})\fright).
\end{align*}

Finally, we choose the same segmentation for both waveguides such that the $z$ component of each segment is the same length $l$. The $z$ center of a segment $Z_j$ shall be at $jl$ and similarly for $S_i$. This means that the straight segment lengths are trivially given by 
$\abs{Z_j} = l$ and
the length of a corresponding helical segment is $\abs{S_i} = l D/P$.
Given these parameters, the height distance between segments $S_i$ and $Z_j$ is $d_z = (i -j)l$.

\subsubsection{Tilted helix axis}
To incorporate the relative geometry of the two waveguides, we tilt the helix axis relative to the straight axis. We introduce a second set of Euler angles $\angleA$, $\angleB$, $\angleC$ analogous to $\alpha$, $\beta$, $\gamma$, which specify the rotation of the helical system. 
Note that $\angleC$ is largely irrelevant due to the symmetry of the helix.
We insert these three new angles into \eqref{SI:eq:rotmatrix} to obtain a secondary rotation matrix which is then applied to the helix parametrization and to its associated Frenet--Serret vectors. 
We then use the latter to compute the Euler angles $\alpha$, $\beta$, $\gamma$ using \eqref{SI:eq:Eulerangles}.
The segment lengths $\abs{S_i}$ and $\abs{Z_j}$ are unchanged by this transformation.

The angle $\theta$ that is varied in the main text is the angle $\angleB$ given here, as it represents the aperture between the helix axis and the straight axis (cf.~Fig.~\ref{SI:fig:geometry}(c)).

\subsection{Results}
\begin{figure}
	\def\svgwidth{220pt}
	\centering
    \includegraphics[width=0.6\textwidth]{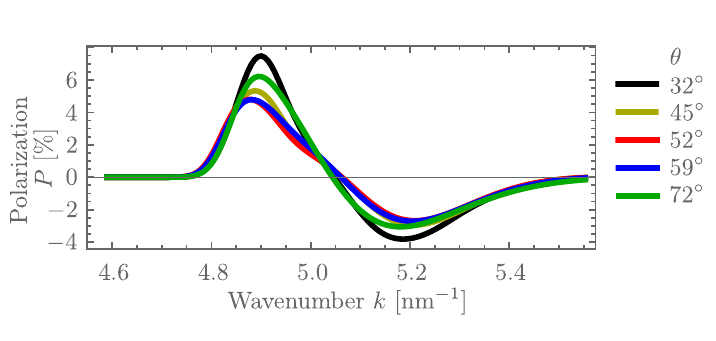}
	\caption{Spin polarization after a single scattering event plotted as a function of the input wavenumber $k$ for various angles $\theta$ between straight waveguide and helix axis for $\hbar\alpha=\SI{10}{\meV\nm}$. Further parameters are stated in the text.}
	\label{SI:Fig:OL_Plot}
\end{figure}
We employ eqs~\ref{SI:eq:I_general} and \ref{SI:eq:Igamma0} to compute coupling matrices which are then inserted into the numerical method described in sec.~\ref{SI:sec:numerics} to obtain polarization dynamics, which we compare to the polarization dynamics found using the simpler diagonal coupling matrix $V(s,q)$ of 
\maneqref{3} in the main text.
As before, we choose the helix parameters $R = \SI{0.7}{\nm}$ and $2\pi P = \SI{3.4}{\nm}$ with $l = 2\pi D/1000 \approx \SI{0.056}{\nm}$. For the remaining parameters, we choose a helix-cross section width $\sigma_n = \SI{0.2}{\nm}$ (small enough to achieve negligible density at the next turn) and a helix position shift of $(x_0, y_0) = (-R, 0)$ such that the electron traveling on the straight waveguide passes exactly through the helical waveguide at the origin. The straight cross-section is chosen as $\sigma_x = \SI{0.5}{\nm}$ which is wider than the helix cross-section but still sufficiently narrow to produce results comparable to those of the simpler $V(s,q)$ model.
Finally, since $\angleA$ and $\angleC$ have a lesser impact on the relative geometry, we keep both of these angles fixed at $A = B = 0$ while varying only $\angleB$, which is named $\theta$ in the main text.
Fig.~\ref{SI:Fig:OL_Plot} shows the average spin-polarization after a single scattering event as a function of the input wavenumber for various angles $\theta$ between straight waveguide and helix axis. The red line ($\theta=\SI{52}{\degree}$) corresponds to a setup where the tangent vectors of straight and helical waveguides are parallel at their point of least separation. As can be seen from the figure, the maximum spin polarization ranges from \SIrange{4}{7}{\percent} for $\theta$ ranging from \SIrange{32}{72}{\degree} while the form of the curves in Fig.~\ref{SI:Fig:OL_Plot} is only slightly affected by the angle. We conclude that the results discussed in the main text are robust against static disorder.

\section{Perturbation theory} \label{SI:sec:perturbation}
We compute first-order transmission and polarization probabilities for the Gaussian coupling model using standard time-dependent perturbation theory. We treat the Hamiltonian of two uncoupled waveguides (helix and free, i.e., 
\maneqref{1}
of the main text and a standard free-particle Hamiltonian, respectively) as the known Hamiltonian $H_0$ and the Gaussian coupling $V$ between the two as the perturbation. Then the first-order time-dependent transition probability at time $t$ from an initial state that is completely localized on the free waveguide at time $t = 0$ to a helix eigenstate $\ket{k,\pm}$ is given by
\begin{equation}
    \abs{\braket{k_\text{out}, \pm}{\psi(t)}}^2 = \abs{\frac{-i}{\hbar} \int_0^t e^{i E(k_\text{out},\pm) t'/\hbar} \mel{k_\text{out}, \pm}{V e^{-iH_0 t'/\hbar}}{\psi_i^\sigma(0)}
 \dd{t'}}^2. \label{SI:eq:PT_basic}
\end{equation}
We use the initial state $\ket{\psi_i^\sigma(0)}$ given in 
\maneqref{5}
with $\sigma =\ \uparrow$ (see below for $\down$).
In $k$ space, this state is given by:
\begin{equation}
\ket{\psi_i^\uparrow(0)} = \left[ \left(\zeta^2/\pi\right)^{\frac{1}{4}} \int_\mathbb{R} e^{-\frac{\zeta^2}{2}\left(k - k_0\right)^2 - ik q_i} \ket{k} \dd{k} \right] \otimes \up \label{SI:eq:gauss_E}.
\end{equation}
Of central importance to the transition probability is the matrix element
\[\mel{k_\text{out}, \pm}{V}{k, \uparrow} = \frac{A_{\pm}^\ast}{2\pi\sqrt{1 + \abs{A_{\pm}}^2}} \int_{\mathbb{R}^2} e^{-is( k_\text{out} - \frac{1}{2D})}  V(s,q) e^{iqk} \dd{s} \dd{q}.\]
We see that this is essentially a Fourier transform of the coupling $V$. Using a Gaussian $V(s,q)$
as in \maneqref{3}:
\begin{align*}
\mel{k_\text{out}, \pm}{V}{k, \uparrow} 
    &= \frac{A_{\pm}^\ast\, V_0 \sqrt{\xi^2} e^{-i\tilde{\phi}}}{\sqrt{2\pi \left(1 + \abs{A_{\pm}}^2\right)}}
    \exp(-\frac{\left(\Delta k\right)^2 \xi^2}{2} - i q_0 \Delta k) \qq{where} \Delta k \coloneqq k_\text{out} - \frac{1}{2D} - k.
\end{align*}
Here, $\tilde{\phi} = \left(k_\text{out} - \frac{1}{2D}\right)\left(s_0 - q_0\right)$ is a $k$-independent phase.
Inserting this matrix element and \eqref{SI:eq:gauss_E} into \eqref{SI:eq:PT_basic}, we get
\begin{equation}
\abs{\braket{k_\text{out}, \pm}{\psi_i^\uparrow(t)}}^2 = 
    \frac{ \abs{A_{\pm}}^2 {V_0}^2 \zeta \xi^2}{2\pi \hbar^2 \sqrt{\pi}\left(1+\abs{A_{\pm}}^2\right)}      \abs{\int_\mathbb{R} 
        e^{-\frac{\xi^2}{2} \left(\Delta k\right)^2} e^{-\frac{\zeta^2 \left(k - k_0\right)^2}{2} - ik (q_i - q_0)} 
            \left[ \int_0^t e^{-i(E_{k} - E(k_\text{out},\pm))t'/\hbar} \dd{t'} \right] 
        \dd{k}}^2.
        \label{SI:eq:SOCallt}
\end{equation}
We define a Bohr frequency:
\begin{equation*}
\omega = \frac{E_{k} - E(k_\text{out},\pm)}{\hbar} = \frac{\hbar k^2}{2m_e} - \frac{1}{\hbar}\left(\varepsilon D^2 {k_\text{out}}^2 \pm k_\text{out} D\Gamma - \frac{\varepsilon_N}{2} \cos{\beta} + \frac{\varepsilon}{4} + \Delta V\right).
\end{equation*}
Using this frequency $\omega$ and the fact that
\[ 
\int_0^t e^{-i \omega t'} \dd{t'} = te^{-it\omega/2} \sinc\fleft(\frac{\omega t}{2}\fright),
\]
we get
\begin{multline*}
\abs{\braket{k_\text{out}, \pm}{\psi_i^\uparrow(t)}}^2 = \frac{ t^2 \abs{A_\pm}^2 {V_0}^2 \zeta \xi^2 }{2\pi \hbar^2 \sqrt{\pi}\left(1+\abs{A_{\pm}}^2\right)} \\
    \times \abs{\int_\mathbb{R} \exp(-\frac{\xi^2\left(k_\text{out} - k - \frac{1}{2D}\right)^2}{2} - \frac{\zeta^2 \left(k - k_0\right)^2}{2} - i \left(k (q_i - q_0) + \frac{\omega t}{2}\right)) \sinc\fleft(\frac{\omega t}{2}\fright) \dd{k}}^2.
    \label{SI:eq:probSOC}
\end{multline*}
On the other hand, for a $\down$-polarized initial state, we follow the same procedure to obtain
\begin{multline*}
\abs{\braket{k_\text{out}, \pm}{\psi_i^\downarrow(t)}}^2 = \frac{ t^2 {V_0}^2 \zeta \xi^2 }{2\pi \hbar^2 \sqrt{\pi }\left(1+\abs{A_{\pm}}^2\right)} \\
    \times \abs{\int_\mathbb{R} \exp(-\frac{\xi^2\left(k_\text{out} - k + \frac{1}{2D}\right)^2}{2} - \frac{\zeta^2 \left(k - k_0\right)^2}{2} - i \left(k (q_i - q_0) + \frac{\omega t}{2}\right)) \sinc\fleft(\frac{\omega t}{2}\fright) \dd{k}}^2,
\end{multline*}
which is identical bar the missing $\abs{A_{\pm}}^2$ in the prefactor and the different sign of $\frac{1}{2D}$ in the exponential (which is precisely the difference in $\Delta k$ due to the $\uparrow$ and $\downarrow$ states).

The preceding two equations are the exact first-order transition probabilities for finite times $t$ and can be numerically integrated to obtain time-resolved dynamics. However, since we are mostly interested in the \enquote{final} states that remain after the wavepackets have fully left the interaction region, we can further investigate the limit $t \to \infty$. To further simplify the calculations, we also take the lower limit of the time integrals to be $ -\infty$ instead of 0 (note that these two yield virtually indistinguishable results if $q_i$ is sufficiently smaller than $q_0$ and $p_0$ is positive and reasonably small, as the wavepacket then undergoes only trivial dynamics in the time interval $(-\infty,0)$).
In this case, instead of the above sinc function, we can use
\[ 
\int_{-\infty}^{\infty} e^{-i \omega t'} \dd{t'} = 2 \pi \delta\fleft(\omega\fright).
\]
Before we carry on with the explicit evaluation of the integrals, we would like to point out the physically intuitive form that we are currently presented with (for $\up$-polarized initial states):
\[
\lim_{t\to\infty} \abs{\braket{k_\text{out}, \pm}{\psi_i^\uparrow(t)}}^2 \propto 
\abs{\int_\mathbb{R} \exp(-\frac{\xi^2}{2} \left(k_\text{out} - \frac{1}{2D} - k\right)^2 -\frac{\zeta^2 \left(k - k_0\right)^2}{2} - ik (q_i - q_0)) \delta\fleft(\omega\fright) \dd{k}}^2,
\]
which is exactly 
\maneqref{6}
of the main text.
The $\delta$ ensures (rigorous) conservation of energy. The two quadratic exponential (and therefore Gaussian) terms act as \enquote{filters} for the transmission: The first ensures weak conservation of momentum, with stronger enforcement for larger $\xi$ (wider interaction regions), whereas the second term corresponds to the initial distribution of momenta around the average momentum $p_0$, i.e., states that are initially weakly occupied will remain weakly occupied. The linear, imaginary term acts carries phase information that is responsible for interference effects. As noted in the main text, this can also be read as a convolution of a Gaussian with the incoming wavefunction in $k$ space.

Continuing the explicit evaluation of the integrals, we use the definition of $\omega$ and the rules for composition of a $\delta$ distribution with a function to find
\[\delta\fleft(\omega\fright) = \sqrt{\frac{m_e}{2 E(k_\text{out},\pm)}} \left[\delta\fleft(k - \frac{\sqrt{2m_e E(k_\text{out},\pm)}}{\hbar}\fright) + \delta\fleft(k + \frac{\sqrt{2m_e E(k_\text{out},\pm)}}{\hbar}\fright)\right].\]
Note that since $k$ is only integrated over the reals, $E_{k}$ is positive, and thus $\omega = 0$ is only fulfilled if $E(k_\text{out},\pm)> 0$. For $E(k_\text{out},\pm) < 0$, the $\delta$ term is zero, which we symbolize using the Heaviside function $\theta\fleft\{E(k_\text{out},\pm)\fright\}$.
Inserting this into \eqref{SI:eq:SOCallt} and using $k_E \coloneqq \frac{\sqrt{2m_e E(k_\text{out},\pm)}}{\hbar}$:
\begin{multline*}
\lim_{t\to\infty} \abs{\braket{k_\text{out}, \pm}{\psi_i^\uparrow(t)}}^2 = \frac{4 \sqrt{\pi} \abs{A_{\pm}}^2 {V_0}^2 \zeta \xi^2 m_e}{\hbar^2 \left(1+\abs{A_{\pm}}^2\right) E(k_\text{out},\pm)}   \theta\fleft\{E(k_\text{out},\pm)\fright\}   
\\
\times 
\exp(-\xi^2 \left[\left(k_\text{out} - \frac{1}{2D}\right)^2 + {k_E}^2\right] - \zeta^2 \left({k_E}^2 + {k_0}^2\right))\
\abs{\cosh\fleft\{k_E 
\left[\xi^2 \left(k_\text{out} -\frac{1}{2D}\right) + \zeta^2 k_0 - i (q_i - q_0)\right]
\fright\}
}^2.
\end{multline*}
As before, for initially $\down$-polarized states, we need to remove $\abs{A_{\pm}}^2$ from the prefactor and flip the sign of every $\frac{1}{2D}$, which changes the weighting of the spin components and shifts the momenta in the opposite direction, respectively.

We finally obtain the total first-order transmission rates of $\up$ and $\down$ initial states to the helix by integrating over the individual transmission probabilities:
\[T_\sigma =  \lim_{t\to\infty} \int_{\mathbb{R}} \left[\abs{\braket{k_\text{out}, +}{\psi_i^\sigma(t)}}^2 + \abs{\braket{k_\text{out}, -}{\psi_i^\sigma(t)}}^2\right] \dd{k_\text{out}} .\]
Then the polarization of the free-waveguide wavepacket after it has left the interaction region can be approximated, to first order, as
\[\mathcal{P}_1 \approx \frac{\left(1 - T_\uparrow\right) - \left(1 - T_\downarrow\right) }{\left(1 - T_\uparrow\right) + \left(1 - T_\uparrow\right)} = \frac{T_\downarrow - T_\uparrow }{2 - T_\uparrow - T_\uparrow} .\]

\bibliography{bibliography}